\pdfoutput=1 
\documentclass{JINST}
\usepackage{amsmath}

\newcommand{\geant}{GEANT4}

\newcommand{\inch}{''}
\newcommand{\hydrogen}{\mbox{$^{1}$H}}

\newcommand{\lith}{\mbox{$^{7}$Li}}

\newcommand{\borten}{\mbox{$^{10}$B}}

\newcommand{\ctwe}{\mbox{$^{12}$C}}
\newcommand{\cfor}{\mbox{$^{14}$C}}

\newcommand{\ar}{\mbox{$^{39}$Ar}}

\newcommand{\pota}{\mbox{$^{40}$K}}

\newcommand{\cobaseven}{\mbox{$^{57}$Co}}
\newcommand{\coba}{\mbox{$^{60}$Co}}

\newcommand{\barium}{\mbox{$^{133}$Ba}}

\newcommand{\cesium}{\mbox{$^{137}$Cs}}
\newcommand{\thal}{\mbox{$^{208}$Tl}}

\newcommand{\bifo}{\mbox{$^{214}$Bi}}
\newcommand{\pofo}{\mbox{$^{214}$Po}}

\newcommand{\radon}{\mbox{$^{222}$Rn}}

\newcommand{\tho}{\mbox{$^{232}$Th}}

\newcommand{\urafive}{\mbox{$^{235}$U}}
\newcommand{\ura}{\mbox{$^{238}$U}}

\newcommand{\AmBe}{\mbox{$^{241}$AmBe}}
\newcommand{\AmC}{\mbox{$^{241}$Am$^{13}$C}}
\newcommand{\alphan}{\mbox{($\alpha$,\,n)}}

\newcommand{\dsf}{\mbox{DarkSide-50}}

\newcommand{\ctf}{\mbox{CTF}}
\newcommand{\lsv}{\mbox{LSV}}
\newcommand{\wcd}{\mbox{WCV}}

\newcommand{\tpc}{\mbox{TPC}}
\newcommand{\pmt}{\mbox{PMT}}
\newcommand{\pmts}{\mbox{PMTs}}

\newcommand{\tmb}{\mbox{TMB}}
\newcommand{\pc}{\mbox{PC}}
\newcommand{\pctmb}{\mbox{\pc-\tmb}}
\newcommand{\ppo}{\mbox{PPO}}

\newcommand{\aar}{\mbox{AAr}}
\newcommand{\lar}{\mbox{LAr}}
\newcommand{\uar}{\mbox{UAr}}

\newcommand{\gr}{\mbox{$\gamma$-ray}}
\newcommand{\grs}{\mbox{$\gamma$-rays}}

\newcommand{\lsvdiameter}{$4.0$\,\mbox{m}}

\newcommand{\lsvpmtnum}{\mbox{110}}

\newcommand{\lsvpmt}{\mbox{Hamamatsu R5912 LRI}}
\newcommand{\lsvpmtsize}{{8}\,\mbox{\inch}}
\newcommand{\lsvpmtqe}{37\%}
\newcommand{\lsvpmtwave}{{408}\,\mbox{nm}}

\newcommand{\lsvcforratehigh}{$\sim$150\,kBq}            
\newcommand{\lsvcforratelow}{0.25\,$\pm$\,0.03\,kBq}             

\newcommand{\lsvppoconcorig}{2.5\,g$/$L}   
\newcommand{\lsvppoconcint}{0.7\,g$/$L}      
\newcommand{\lsvppoconcnow}{$1.4$\,g$/$L}   
\newcommand{\ctfpmtnum}{\mbox{80}}

\newcommand{\ctfpmt}{\mbox{ETL 9351}}
\newcommand{\ctfpmtsize}{8\,\inch}

\newcommand{\ctfheight}{{10}\,\mbox{m}}
\newcommand{\ctfdiameter}{{11}\,\mbox{m}}

\newcommand{\phaseone}{Phase-I}
\newcommand{\phasetwo}{Phase-II}
\newcommand{\aftp}{after-pulse}

\newcommand{\enbortenexcitedgamma}{478\,\mbox{keV}}
\newcommand{\brbortenground}{$6.4${\%}}

\newcommand{\brbortenexcited}{$93.6${\%}}

\newcommand{\lngsmuflux}{1.1\,muons/(m$^2\cdot$hr)}
\newcommand{\lngsdepthnounits}{3800} 
\newcommand{\lngsdepth}{\lngsdepthnounits\,\mbox{m.w.e.}}  
\newcommand{\dsfexpo}{{(1422\,$\pm$\,67)}\,\mbox{kg-day}}
\newcommand{\pxie}{\mbox{PXIe}}

\title{The veto system of the DarkSide-50 experiment}
\author{
	P.~Agnes$^a$,
	L.~Agostino$^b$,
	I.~F.~M.~Albuquerque$^{c,d}$,
	T.~Alexander$^{e,f}$,
	A.~K.~Alton$^g$,
	K.~Arisaka$^h$,
	H.~O.~Back$^{c,i}$,
	B.~Baldin$^f$,
	K.~Biery$^f$,
	G.~Bonfini$^j$
% ...
	\\
	(The DarkSide Collaboration)
	\\
\llap{$^a$}APC, Universit\'e Paris Diderot, CNRS/IN2P3, CEA/Irfu, Obs de Paris, Sorbonne Paris Cit\'e, 75205 Paris, France\\
\llap{$^b$}LPNHE Paris, Universit\'e Pierre et Marie Curie, Universit\'e Paris Diderot, CNRS/IN2P3, Paris 75252, France\\
\llap{$^c$}Department of Physics, Princeton University, Princeton, NJ 08544, USA\\
\llap{$^d$}Instituto de F\'isica, Universitade de S\~ao Paulo, S\~ao Paulo 05508-090, Brazil\\
\llap{$^e$}Amherst Center for Fundamental Interactions and Department of Physics, University of Massachusetts, Amherst, MA 01003, USA\\
\llap{$^f$}Fermi National Accelerator Laboratory, Batavia, IL 60510, USA\\
\llap{$^g$}Department of Physics, Augustana University, Sioux Falls, SD 57197, USA\\
\llap{$^h$}Department of Physics and Astronomy, University of California, Los Angeles, CA 90095, USA\\
\llap{$^i$}Pacific Northwest National Laboratory, Richland, WA 99354, USA\\
\llap{$^j$}Laboratori Nazionali del Gran Sasso, Assergi AQ 67010, Italy\\
\llap{$^k$}Gran Sasso Science Institute, L'Aquila 67100, Italy\\
\llap{$^l$}Department of Physics, Universit\`a degli Studi, Genova 16146, Italy\\
\llap{$^m$}Istituto Nazionale di Fisica Nucleare, Sezione di Genova, Genova 16146, Italy\\
\llap{$^n$}Istituto Nazionale di Fisica Nucleare, Sezione di Milano, Milano 20133, Italy\\
\llap{$^o$}Istituto Nazionale di Fisica Nucleare, Sezione di Roma Tre, Roma 00146, Italy\\
\llap{$^p$}Department of Physics and Mathematics, Universit\`a degli Studi Roma Tre, Roma 00146, Italy\\
\llap{$^q$}Department of Physics, Universit\`a degli Studi, Cagliari 09042, Italy\\
\llap{$^r$}Istituto Nazionale di Fisica Nucleare, Sezione di Cagliari, Cagliari 09042, Italy\\
\llap{$^s$}Department of Physics, University of Houston, Houston, TX 77204, USA\\
\llap{$^t$}Istituto Nazionale di Fisica Nucleare, Sezione di Napoli, Napoli 80126, Italy\\
\llap{$^u$}Department of Physics, Universit\`a degli Studi Federico II, Napoli 80126, Italy\\
\llap{$^v$}Department of Physics, Virginia Tech, Blacksburg, VA 24061, USA\\
\llap{$^w$}Skobeltsyn Institute of Nuclear Physics, Lomonosov Moscow State University, Moscow 119991, Russia\\
\llap{$^x$}Department of Physics, Universit\`a degli Studi, Milano 20133, Italy\\
\llap{$^y$}St. Petersburg Nuclear Physics Institute NRC Kurchatov Institute, Gatchina 188350, Russia\\
\llap{$^z$}Department of Physics and Astronomy, University of Hawai'i, Honolulu, HI 96822, HI\\
\llap{$^{aa}$}Joint Institute for Nuclear Research, Dubna 141980, Russia\\
\llap{$^{bb}$}Department of Physics, Temple University, Philadelphia, PA 19122, USA\\
\llap{$^{cc}$}Kavli Institute, Enrico Fermi Institute and Dept. of Physics, University of Chicago, Chicago, IL 60637, USA\\
\llap{$^{dd}$}Institute for High Energy Physics, Beijing 100049, China\\
\llap{$^{ee}$}Laboratorio Subterr\'aneo de Canfranc, Canfranc Estaci\'on E-22880, Spain\\
\llap{$^{ff}$}IPHC,19 Universit\'e de Strasbourg, CNRS/IN2P3, Strasbourg 67037, France\\
\llap{$^{gg}$}School of Natural Sciences, Black Hills State University, Spearfish, SD 57799, USA\\
\llap{$^{hh}$}Institute for National Research, National Academy of Sciences of Ukraine, Kiev 03680, Ukraine\\
\llap{$^{ii}$}Radiation Physics Laboratory, Belgorod National Research University, Belgorod 308007, Russia\\
\llap{$^{jj}$}SLAC National Accelerator Laboratory, Menlo Park, CA 94025, USA\\
\llap{$^{kk}$}National Research Centre Kurchatov Institute, Moscow 123182, Russia\\
\llap{$^{ll}$}National Research Nuclear University MEPhI, Moscow 115409, Russia\\
\llap{$^{mm}$}Department of Chemistry, Biology and Biotechnology, Universit\`a degli Studi, Perugia 06123, Italy\\
\llap{$^{nn}$}Istituto Nazionale di Fisica Nucleare, Sezione di Perugia, Perugia 06123, Italy\\
\llap{$^{oo}$}Department of Physics, University of California, Davis, CA 95616, USA\\
\llap{$^{pp}$}Smoluchowski Institute of Physics, Jagiellonian University, Krakow 30059, Poland\\
\llap{$^{qq}$}Lawrence Livermore National Laboratory, Livermore, CA 94550, USA\\
\llap{$^{rr}$}Institute of Physics Gleb Wataghin  Universidade Estadual de Campinas, S\~ao Paulo 13083-859, Brazil\\
\llap{$^{ss}$}School of Physics, University of Chinese Academy of Sciences, Beijing 100049, China\\
}

\abstract{
Nuclear recoil events produced by neutron scatters form one of the most important classes of background in WIMP direct detection experiments, as they may produce nuclear recoils that look exactly like WIMP interactions.
In DarkSide-50, we both actively suppress and measure the rate of neutron-induced background events using our \emph{neutron veto}, composed of a boron-loaded liquid scintillator detector within a water Cherenkov detector. 
This paper is devoted to the description of the neutron veto system of DarkSide-50, including the detector structure, the fundamentals of event reconstruction and data analysis, and basic performance parameters.
}


\begin{document}

\section{Introduction}
\label{sec:introduction}

\subsection{\dsf\ and its neutron veto}

Experiments attempting to directly detect dark matter in the form of Weakly Interacting Massive Particles (WIMPs) through WIMP-nucleon scattering must be carefully designed to reduce neutron-, $\beta$-, and \gr-induced backgrounds~\cite{goodman_detectability_1985}.
In the \dsf\ experiment at Gran Sasso National Laboratory (LNGS),
electron recoils can be rejected effectively by the Liquid Argon Time Projection Chamber (\lar\ \tpc),
leaving nuclear recoil events as the dominant background~\cite{Agnes:2015gu, Agnes:2015_uar}.
Nuclear recoils can come either from $\alpha$ decays on the inner surfaces of the detector, in which the $\alpha$ goes into the surface and the recoiling nucleus scintillates in the \lar, or from neutrons scattering on argon nuclei.
The former backgrounds can be eliminated by fiducializing the \lar\ to remove events near the surfaces.
This leaves nuclear recoils induced by single neutron scatters as one of the most troublesome backgrounds because they are indistinguishable from WIMP interactions.

\begin{figure}[tb]
 \centering
 \includegraphics[width=.8\linewidth]{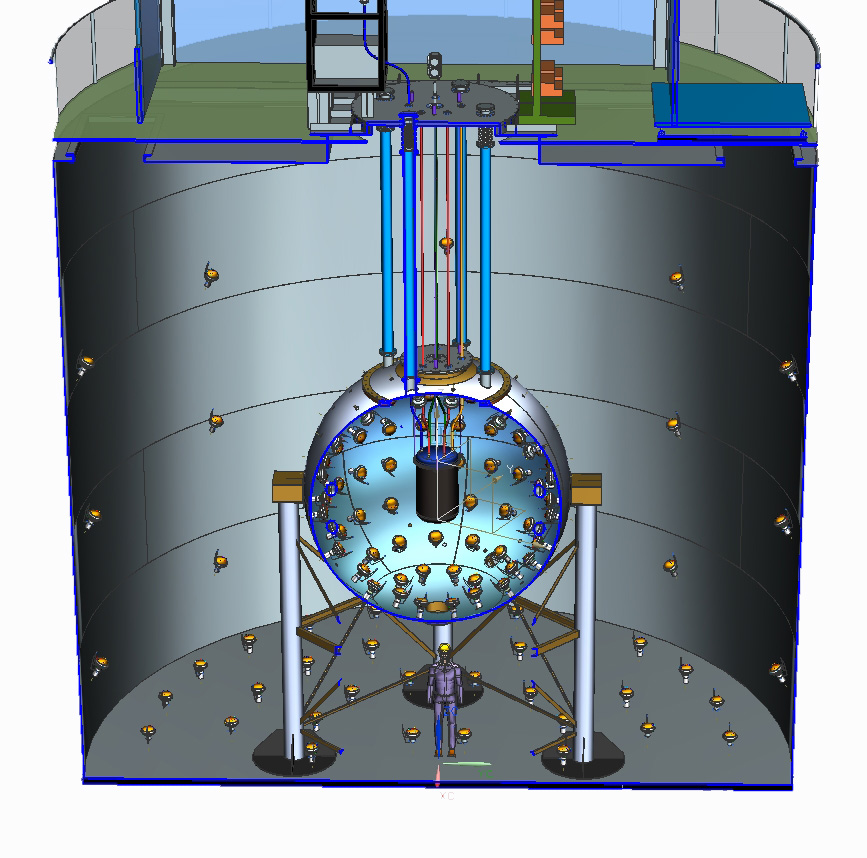}
 \caption{Schematic drawing of the \dsf\ detector at LNGS. The outermost cylindrical tank is the \wcd, the sphere is the \lsv, and the inner cylinder is the \lar\ \tpc\ cryostat. The radon-free clean room (CRH) is on top of the \wcd.}
 \label{fig:ds50}
\end{figure}

The limited size of the \lar\ \tpc\ of \dsf\ does not allow a fiducial volume well-shielded from neutron-induced backgrounds. 
Passive shielding could reduce the level of neutron-induced nuclear recoils, but it is challenging for an experiment with only passive shielding to conclusively measure the background levels. This makes the interpretation of a few observed recoil events as a WIMP signal problematic. 
Additionally, external passive shielding does not protect against radiation from the detector components themselves or from high energy (often $\sim$100\,MeV) cosmogenic neutrons, which may penetrate deeply through most shielding~\cite{formaggio_backgrounds_2004}. 
It should also be noted that the comparatively shallow depth of LNGS (\lngsdepthnounits\ meters water equivalent, m.w.e.) means that the rate of cosmogenic neutron backgrounds is higher than in very deep sites, making an additional active layer of shielding crucial to the detector design.

A better method of neutron suppression is the use of an active neutron detector (the \emph{neutron veto}) in which the neutrons from both internal and external sources are detected with very high efficiency, and the corresponding recoil events induced by neutrons in the argon are thus identified and rejected.
In addition to removing neutron backgrounds, an active veto also provides \emph{in situ} measurements of the actual neutron background in the experiment, allowing background models to be compared to the data seen in the veto. For example, decay rates of various isotopes in the uranium and thorium decay chains can be measured by comparing their signals in the veto to those expected from Monte Carlo, and these rates can be used to compute the expected fission and \alphan\ neutron rates. This comparison allows much more reliable predictions of the number of neutron-induced recoils that are not vetoed.
In \dsf\ we actively veto radiogenic and cosmogenic neutron backgrounds with a water Cherenkov- and boron-loaded liquid scintillator-based  veto system. This veto system serves both as shielding and as a tag for radiogenic and cosmogenic neutrons, cosmic muons, and \grs.
The key concepts and the design goals of this kind of active neutron veto system for dark matter searches
are described in~\cite{Wright:2011ig}.

The \dsf\ experimental apparatus consist of three nested detectors (see figure~\ref{fig:ds50}).
From the center outward, these detectors are the 50-kg \lar\ \tpc, the WIMP-sensitive volume of the experiment,
the 30-tonnes Liquid Scintillator Veto (\lsv), and the 1000-tonnes Water Cherenkov Veto (\wcd).
We refer to the \lsv\ and \wcd\ detectors together as  the \emph{veto system} of \dsf.
The \dsf\ detectors are located in Hall C of LNGS, in close proximity to and sharing many facilities with the Borexino experiment~\cite{Bellini:2014ke, Alimonti:2009dd}.

The \dsf\ collaboration is conducting a WIMP search with the \lar\ \tpc\ filled with argon derived from underground sources (called Underground Argon or \uar), which has a lower concentration of the radioactive \ar\ than is present in Atmospheric Argon (\aar).
A first run of \dsf\ with a \dsfexpo\ exposure of \aar\ produced a null result for the dark matter search and zero backgrounds from \ar\ decays~\cite{Agnes:2015gu}.
A total of 16 million background events in the \lar\ \tpc, mostly originating from \ar, were collected.
All but two of the events falling within the WIMP region of interest were rejected using the \lar\ primary-scintillation Pulse Shape Discrimination (PSD). PSD distinguishes
between electron recoil signals ($\beta$ and $\gamma$ decays) from nuclear recoil signals (neutrons or WIMPs).
The two remaining events in the WIMP search region had a signal in coincidence with the veto
and were therefore discarded. 
The first  WIMP search in \dsf\ using \uar\ has been also reported in~\cite{Agnes:2015_uar},
where it is shown that \uar\ is depleted in \ar\ by a factor $(1.4\, \pm\,0.2) \, \times \, 10^3$ relative to \aar.
The combination of the electron recoil background rejection observed in the \aar\ run, and the reduction of \ar\ from the use of \uar\ would allow \dsf\ to be free from \ar\ background for several tens of years.
Neutron recoils are then the primary remaining background that must be addressed in \dsf.

\subsection{Outline of the paper}

This paper describes the neutron veto system of \dsf, the analysis procedures to identify and reject the neutron backgrounds and the system performance.
Both the neutron rejection power and the comparison between the measured and expected neutron background will be subjects of a future paper.

Section~\ref{sec:neutrons} of this paper provides relevant physics background underlying the neutron production and detection mechanisms.
Sections~\ref{sec:lsv} and~\ref{sec:wcd} describe the \lsv\ and  \wcd\ detector respectively, while
section~\ref{sec:reconstruction} summarizes the event reconstruction.
Finally section~\ref{sec:performances} provides an overview of the veto performance, such as the response of the \lsv\ to neutron captures in the various phases of the \dsf\ experiment.

\section{Neutron production and detection mechanisms}
\label{sec:neutrons}
In order to understand the design and efficiency of the neutron veto, it is helpful to first discuss the physics underlying the source of the neutron backgrounds and the possible neutron detection mechanisms.

\subsection{Neutron background sources}

Neutrons that enter the \tpc\ of \dsf\ and interact with the \lar\ are expected to come primarily from four sources.
\begin{itemize}
 \item Fission reactions in the detector materials --- Uranium and thorium contamination in the detector components may fission and produce neutrons.
 \item \alphan\ reactions in the detector materials --- $\alpha$-emitting radioisotopes contaminating the detector components may interact with light nuclei to produce neutrons in the detector materials.
  \item Radioactivity in the environment --- Uranium and thorium in the surrounding rocks can produce fission or \alphan\ neutrons that may then enter the \lar\ \tpc.
 \item Cosmogenic interactions~\cite{formaggio_backgrounds_2004} --- Cosmic ray muons can interact with the detector or surrounding materials to produce high energy (typically around 100\,MeV) neutrons, or they can produce unstable nuclei that will decay and produce lower energy neutrons.
\end{itemize}

Neutrons from the last two sources are  easily eliminated by the veto design: the \wcd\ detects passing muons, and the \lsv\ provides a substantial amount of shielding as well as a visible signal before external neutrons can reach the \lar\ \tpc. Previous simulations have shown that a buffer of at least 1\,m of boron-loaded liquid scintillator can drastically reduce the rate of external neutrons that would otherwise create a dangerous background~\cite{empl_fluka_2014}. Additionally, fission reactions that produce neutrons often generate multiple neutrons and high energy \grs, significantly increasing the probability of multiple coincident interactions in the \lsv\ at the same time that a neutron interacts with the \lar. This leaves \alphan\ neutrons as the most challenging type of neutron to veto, and  much of the design and analysis is targeted around vetoing these neutrons with a  high efficiency.

The rate of neutrons from \alphan\ reactions which undergo a single elastic scatter in the \lar\ \tpc, imparting a nuclear recoil kinetic energy within the WIMP search region, 
determines the neutron veto efficiency needed to run the experiment background-free for a given time.
For example, for an expected rate of 10 neutrons per year giving single scatters in the \lar\ \tpc\ (as roughly estimated from Monte Carlo simulations of \dsf~\cite{Agnes:2015gu}),
a neutron rejection efficiency of 99\% (99.7\%) allows one (three) year of \dsf\ data taking with less than $0.1$ expected nuclear
recoil background from neutrons.

\subsection{Neutron detection mechanisms}\label{sec:neutron_detection}
When a neutron enters the \lsv\ from the \lar\ \tpc, there are two signals that can be used to detect and veto the neutron.

The first signal is the prompt thermalization signal produced by the neutron slowing down in the \lsv. 
Neutrons lose energy by scattering off the nuclei in the scintillator, mostly on hydrogen and carbon, with most of its energy being lost to the former.
While measurements of the scintillator's	 response to nuclear recoils are awaiting future calibration campaigns with an Americium-Carbon neutron source~\cite{liu_neutron_2015}, measurements made with a similar scintillator in~\cite{Hong:2002fn, GrauMalonda:1999ex} indicate 
that scintillation from proton recoils is quenched to about 5--10\% of that from electron recoils
of the same kinetic energy, while carbon recoil scintillation is quenched to about 1--5\%.
Monte Carlo simulations have shown that the thermalization signal follows the signal in the \lar\ \tpc\ very quickly, with nearly all of the energy deposited into the scintillator within $\sim$100\,ns of the recoil in the \tpc.

The second signal is the delayed signal from  the neutron capture. The thermalized neutron can capture on various isotopes in the scintillator cocktail, typically on a time scale on the order of 1--100 $\mu$s. For example, in the \dsf\ \lsv\, captures may occur on \borten, \hydrogen, or \ctwe, in a delayed coincidence typically within several tens of microseconds of the neutron scatter in the \lar\ \tpc. These capture reactions are discussed in more detail in sections~\ref{sec:scint} and \ref{sec:reco-ambe}. An important feature of this signal is that  it is independent of the energy of the incoming neutron. This means that a neutron that thermalizes in the \tpc\ detector components or has too low energy to produce a detectable prompt signal in the \lsv\ may still produce a detectable signal when captured.

In addition to detecting neutrons through the above two primary modes, it is also possible to indirectly veto neutron events in the \lar\ \tpc. If a neutron scatters in the \tpc\ but captures on detector components, this capture reaction may produce a \gr\ that can later be detected in the \lsv\ on a timescale up to 60\,$\mu$s after the scatter in the \tpc, as shown in Monte Carlo simulations~\cite{Wright:2011ig}. 
\dsf\ is designed to maximize the chance of a neutron that scatters in the TPC making it into the \lsv. This is achieved by 
avoiding the use of detector materials with light nuclei --- especially hydrogen --- and materials with high neutron capture cross sections.
The highest probability of neutron capture on the \tpc\ detector material comes from Teflon, where the neutron may capture on $^{19}$F and produce a $6.6$\,MeV \gr~\cite{endf}. It is also possible for neutrons to capture on $^{56}$Fe in the stainless steel cryostat and produce a 7.6\,MeV \gr~\cite{endf}. The thermal capture cross sections of $^{19}$F and $^{56}$Fe are $0.01$ and $2.5$\,barns, respectively~\cite{Wright:2011ig}.

If a neutron is produced by a cosmic ray muon, it is possible to indirectly veto the neutron by detecting the muon in the \wcd. Alternatively, if the muon produces neutrons in the rock surrounding the experiment and does not pass through the \wcd, it may be possible to detect charged products of the electromagnetic shower accompanying the neutron's production in the rocks.

Neutrons that scatter once in the \lar\ \tpc\ and don't produce any signal in either the \lsv\ or the \wcd, directly or indirectly, will not be vetoed and may produce a fake WIMP-like event in \dsf.
Monte Carlo simulations show that only $\sim$0.05\% of radiogenic neutrons produced by the ($\alpha,n$) reaction leave absolutely no signal in the \lsv.

\section{The Liquid Scintillator Veto}
\label{sec:lsv}

\begin{figure}[tb]
 \centering
 \includegraphics[width=.4\linewidth]{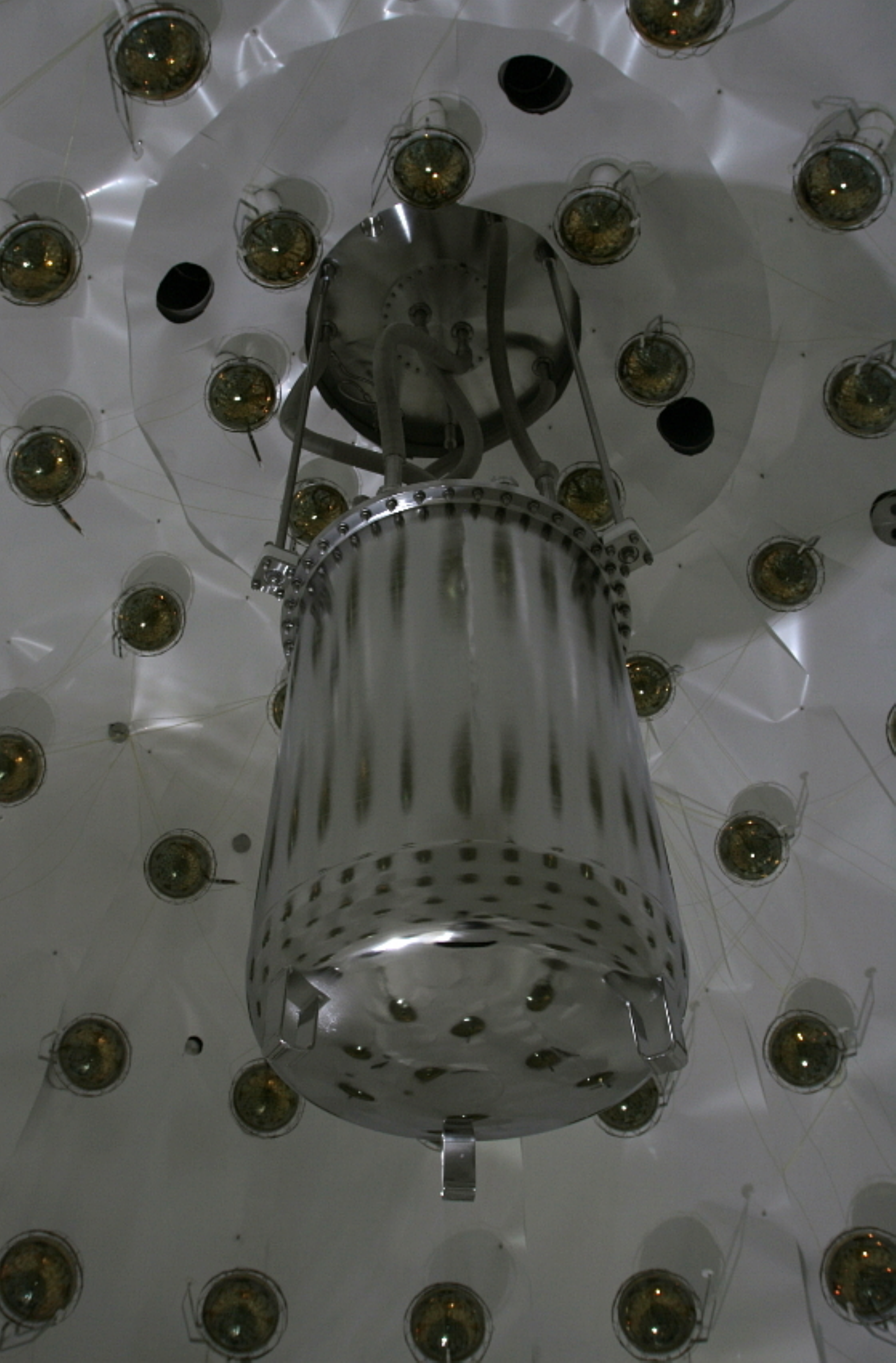}
 \caption{The Liquid Scintillator Veto (\lsv). The picture shows the cryostat of the \dsf\ \lar\ \tpc\ hanging from the top.}
 \label{fig:lsvpic}
\end{figure}

\begin{figure}[tb]
 \centering
 \includegraphics[width=0.8\linewidth]{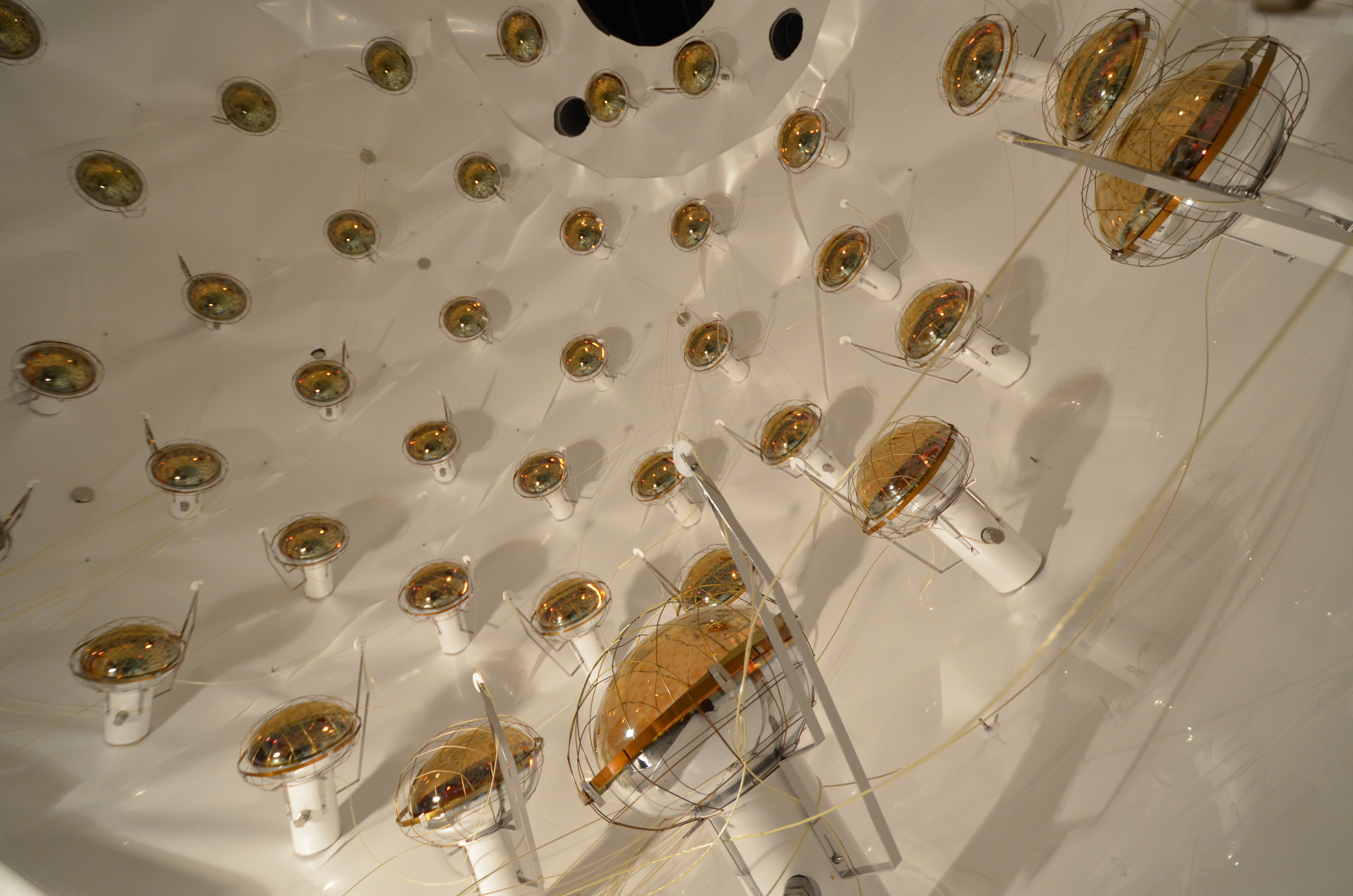}
 \caption{The Liquid Scintillator Veto. The picture shows the internal surface of the \lsv, covered with Lumirror,
 with \pmts\ evenly distributed inside.}
 \label{fig:lsvpic2}
\end{figure}

The \lsv\ is a \lsvdiameter\ diameter stainless steel sphere filled with 30 metric tonnes of boron-loaded liquid scintillator.
The sphere is lined with Lumirror, a reflecting foil used to enhance the light collection efficiency.
An array of \lsvpmtnum\ \lsvpmt\ \lsvpmtsize\ \pmts\ is mounted on the inside surface of the sphere to detect scintillation photons.
Photographs of the inside of the \lsv\ detector  can be seen in figures~\ref{fig:lsvpic} and~\ref{fig:lsvpic2}.

The neutron capture reaction \borten(${\mbox n},\,\alpha$)\lith\ makes the boron-loaded scintillator a very effective neutron veto
because of its large cross section for thermal neutron capture leading to charged products.
An $\alpha$ particle and a \lith\ nucleus are always produced as a result of the neutron capture on \borten. 
Because the $\alpha$ and the \lith\ nucleus are short ranged charged particles, a neutron capture on \borten\ in the \lsv\ scintillator will always deposit visible energy in the detector.

\subsection{Organic liquid scintillator}
\label{sec:scint}

\begin{figure}[tb]
 \centering
 \includegraphics[width=0.8\linewidth]{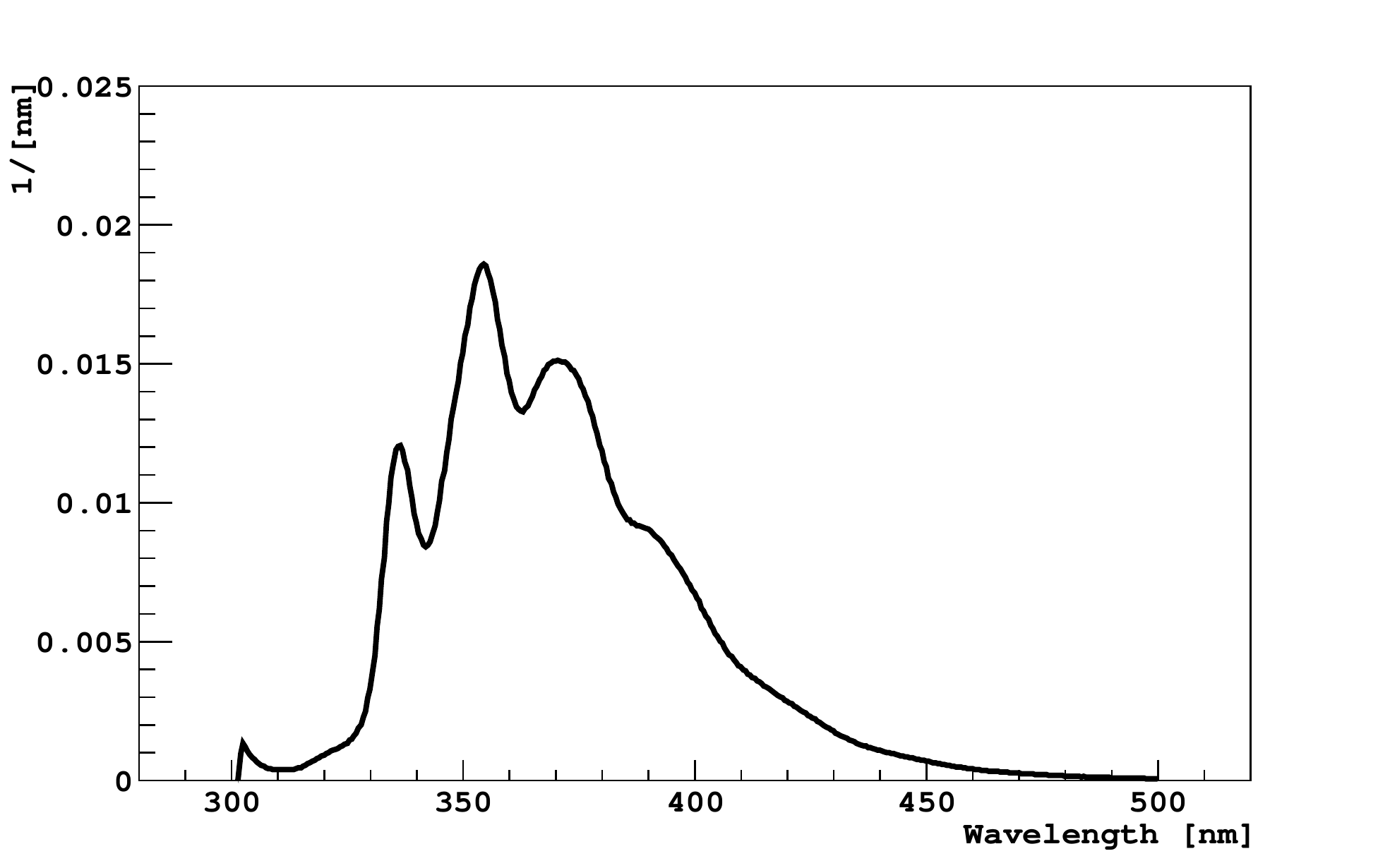}
 \caption{The emission spectrum of PPO~\cite{Berlman}.}
 \label{fig:lsEmission}
\end{figure}

\begin{figure}[tb]
 \centering
 \includegraphics[width=0.8\linewidth]{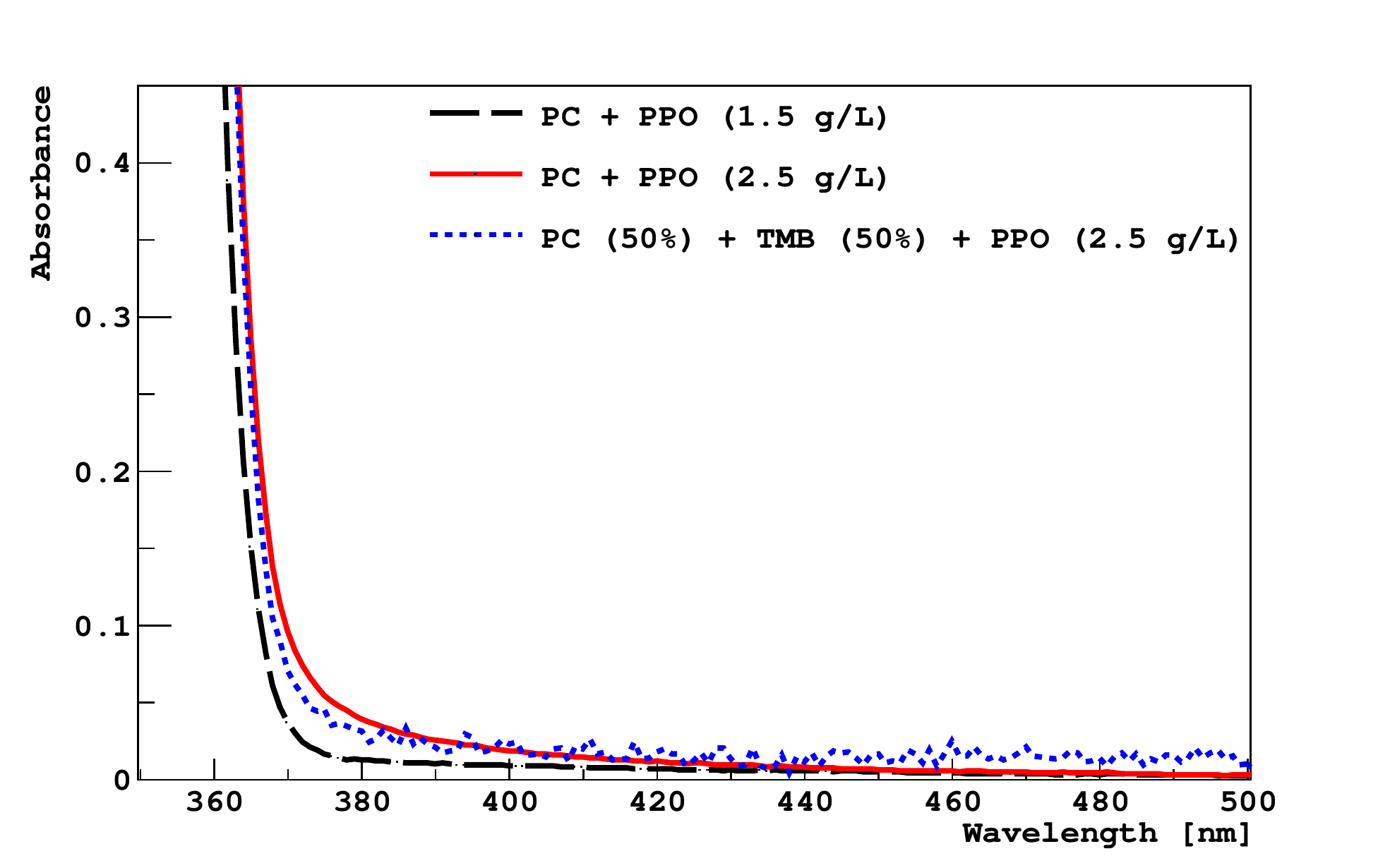}
 \caption{The absorbance of three different sample scintillator cocktails. The absorbance is defined as $\log_{10}(I/I_0)$, where $I_0$ and $I$ are the incoming and outgoing light intensity, respectively, measured in a cell containing a sample of each scintillator. The samples measured were (black, dashed) \pc\ with 1.5\,g/L \ppo\ in a 10\,cm cell, (red, solid) \pc\ with \lsvppoconcorig\, \ppo\ in a 10\,cm cell, and (blue, dotted) 50\% \pc\ and 50\% \tmb\ and \lsvppoconcorig\, \ppo\ in 1\,cm cell. At low absorbances, the blue curve becomes statistically limited in its measurements due to the small size of the cell.}
 \label{fig:lsAbsorbance}
\end{figure}

The boron-loaded liquid scintillator has three primary components: pseudocumene (\pc), trimethyl borate (\tmb), and 2,5-diphenyloxazole (\ppo). 
\pc, C$_6$H$_3$(CH$_3$)$_3$, is the primary scintillator used in the veto and makes up the bulk of the cocktail. 
\tmb, B(OCH$_3$)$_3$, is an organic molecule containing one boron atom. \borten, with a natural abundance of $19.9$\%, has a very high thermal neutron capture cross section of 3837(9)\,barn~\cite{Wright:2011ig}.

When a neutron captures on \borten, two reactions are possible:

\begin{equation}
^{10}\text{B}+n\rightarrow \begin{cases}^7\text{Li (1015 keV)}+\alpha \text{ (1775 keV)} & \mbox{(6.4\%)}\\
^7\text{Li}^* + \alpha \text{ (1471 keV)}, ^7\text{Li}^*\rightarrow ^7\text{Li (839 keV)} + \gamma \text{ (478 keV)} & \mbox{(93.6\%)} \end{cases}
\end{equation}


The decay to the excited state produces a \gr\ that is easily seen as long as it does not escape into the cryostat before depositing energy into the scintillator. 
Energy deposits due to the $\alpha$ and \lith\ nucleus, on the other hand, are always contained in the scintillator, due to their high stopping power and consequently short track length. 
This gives boron a comparative advantage over other loading options such as gadolinium, which only produces high energy \grs\ which may escape the veto without leaving a detectable signal.
However, the light output of $\alpha$ and \lith\ nuclei is highly suppressed due to ionization quenching, causing them to scintillate equivalently to a 50--60\,keV electron (an energy scale referred to as ``keV electron equivalent'', or keVee).
Detecting these decay products therefore requires a high light collection efficiency and low background.
If the detector can very reliably detect these nuclear decay products, it can very efficiently detect neutrons that capture in the veto, regardless of their initial kinetic energy, since the energy of these capture products does not depend on the neutron's initial energy. 

In addition to capturing on \borten, a thermal neutron may also capture on \hydrogen\ or \ctwe\ through the following reactions
\begin{gather}
\begin{aligned}
  ^{1}\text{H}+n &\rightarrow ^2\text{H} + \gamma \text{ (2223 keV)} \hphantom{1000} I_\gamma/I_\gamma(\text{max}) = 100\% && \sigma = 0.33\text{b}\\
  ^{12}\text{C}+n&\rightarrow 
  \begin{cases} 
  ^{13}\text{C} + \gamma \text{ (3090 keV)} & I_\gamma/I_\gamma(\text{max}) = 100\%\\
  ^{13}\text{C} + \gamma \text{ (4945 keV)} & I_\gamma/I_\gamma(\text{max}) = 67\% \\
  ^{13}\text{C} + \gamma \text{ (1860 keV)} & I_\gamma/I_\gamma(\text{max}) = 57\%
  \end{cases}
   &&\sigma = 0.0034\text{b}
\end{aligned}
\end{gather}

where $\sigma$ is the thermal neutron capture cross section,  $I_\gamma/I_\gamma(\text{max})$ is the intensity of the \gr, relative to the maximum intensity \gr~\cite{endf}. For \ctwe\ only the three dominant \grs\ are shown (notably, \ctwe\ will often produce multiple \grs\ after capturing a neutron).

The wavelength shifter \ppo\ is added in a concentration of a few grams per liter in order to increase the detection efficiency of the \lsv. 
Since energy deposited in the \pc\ can non-radiatively transfer to the \ppo, only a small concentration is needed for all of the light to be efficiently shifted to a longer wavelength, increasing the attenuation length of light in the scintillator. Additionally, since \ppo\ scintillates much faster than \pc, adding \ppo\ can make the light signal faster, allowing for tighter prompt coincidence cuts.
Lastly, the Borexino collaboration has observed that increasing \ppo\ concentration decreases the effects of ionization quenching, which would make it easier to detect the nuclear decay products~\cite{elisei_measurements_1997, chen_quenching_1999}.

The emission spectrum of the \ppo\ is shown in figure~\ref{fig:lsEmission}. 
The absorption lengths of various scintillator cocktails relevant to the \dsf\ liquid scintillator are shown in figure~\ref{fig:lsAbsorbance}. As can be seen from these two plots, the upper half of the PPO emission spectrum has a long attenuation length in the scintillator cocktail, allowing the \lsv\ to reach a high light yield.

The \lsv\ was first filled with boron-loaded scintillator during the first two weeks of October, 2013.
Two different mixtures of \pc, \tmb, and \ppo\ were used in the two WIMP search phases of \dsf~\footnote{\phaseone\ of \dsf\ corresponds also to the WIMP search with \aar~\cite{Agnes:2015gu}, while \phasetwo\ corresponds to the WIMP search with \uar~\cite{Agnes:2015_uar}.}:
\begin{itemize}
 \item \emph{\phaseone} Nov. 2013 -- June 2014: 50\% mass fraction of  \pc, 50\% \tmb, \lsvppoconcorig\ \ppo\
 \item \emph{\phasetwo} Feb. 2015 -- present:  95\% mass fraction of \pc, 5\% \tmb, \lsvppoconcnow\ \ppo\
\end{itemize}

The capture times and relative rate of neutrons capturing on \hydrogen, \borten, and \ctwe\ can be calculated from the cross sections given above and the chemical compositions and concentrations of \pc\ and \tmb. These rates and capture times were calculated and compared with measurements taken during an \AmBe\ source calibration run. The measurements were found to agree with the predictions, as described in section~\ref{sec:reco-ambe}. 
During \phaseone\ of \dsf, with a 50\% \tmb\ concentration, the neutron capture time was $\sim2.2\,\mu$s and $\sim$0.8\% of neutron captures were expected to be on \hydrogen. 
During \phasetwo, with a $\sim$5\% concentration, the neutron capture time is $\sim22\,\mu$s and we expect $\sim$8\% of neutrons to capture on \hydrogen. 
The number of captures on \ctwe\ should be about two orders of magnitude below the \hydrogen\ capture rate. 

During \phaseone, it was found that the \lsv\ exhibited a high rate of \cfor\ decays (\lsvcforratehigh).
The endpoint of \cfor\ beta spectrum is 156\,keV, and so this high rate of \cfor\ decays made the $\alpha$ and \lith\ decay products impossible to be distinguished over the background, severely limiting the rejection power of the veto.
\cfor\ is naturally formed in the atmosphere by the cosmogenic activation of nitrogen and its abundance has been further increased due to nuclear weapons testing. 
This results in trace amounts of \cfor\ ($\sim 10^{-12}$\,g$/$g) being present in all natural carbon samples. 
Upon further investigation, it was discovered that the Dow Chemicals plant located in the Netherlands that produced the \tmb\ used biogenic methanol to derive much of the \tmb, rather than methanol from petroleum. Since biogenic methanol gets much of its carbon from the atmosphere, this \tmb\ had a very high \cfor\ concentration.

Petroleum-dervied methanol tends to have much lower levels of \cfor, because the petroleum used for these processes has spent millions of years shielded from cosmic rays underground, far longer than the 5700 year halflife of \cfor. 
A \cfor$/$\ctwe\ ratio of $\sim 2 \times 10^{-18}$ was measured in the Borexino CTF for petroleum feedstock~\cite{Alimonti:1998hd}. 
After identifying several sources of petroleum-derived \tmb, samples were sent to Lawrence Livermore National Laboratory for analysis by accelerator mass spectroscopy. 
A sample measured to have a \cfor\ concentration below the limit of detectability of 10$^{-3}$ times the modern \cfor\ fraction (equivalent to \cfor$/$\ctwe\, < 10$^{-15}$) was chosen. 
The chosen sample was produced by the Dow Chemicals plant in the US, which used a petroleum-based process. 
\tmb\ from the US Dow plant was then purchased to replace the high \cfor\ \tmb\ at a lower concentration of 5\%, primarily for cost reasons.
The effects of \cfor\ and \tmb\ on the performance of the \lsv\ are discussed in more detail in section~\ref{sec:performances}.

\subsection{Fluid handling system and scintillator purification}
\label{sec:fluidhandling}
\label{sec:lsv_reconstitution}

\begin{figure}[tbp]
\centering
\includegraphics[width=0.9\textwidth]{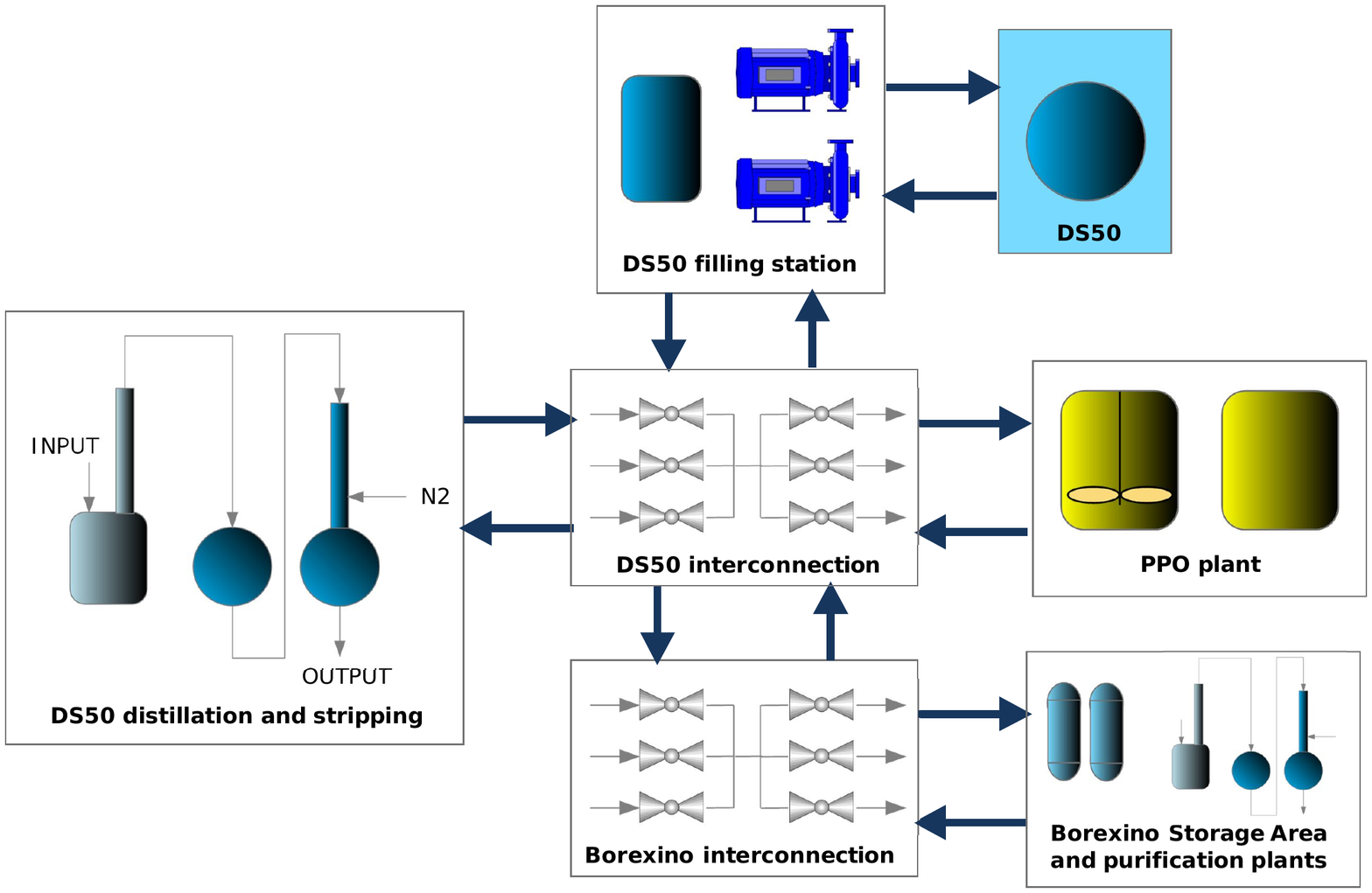}
\caption{Block diagram of the liquid handling systems for the active veto of the \dsf\ experiment.}
\label{fig:plants}
\end{figure}

\begin{table}[tbp]
\centering
\caption{Main parameters used for the scintillator and master solution purification.}
\begin{tabular}{l | c c}
 \hline
  & Scintillator & Master Solution \\
  \hline
  \hline
  Flow & 90\,kg$/$h & 40\,kg/h \\
  Reboiler temperature & 40\,C & 180\,C \\
  Condensing temperature & 20\,C & 20\,C \\
  Reboiler pressure & 55\,mbar & 55\,mbar\\
  Condensing pressure & 50\,mbar & 50\,mbar\\
  Nitrogen flow & 10\,kg$/$h & 10\,kg$/$h\\
  \hline
 \end{tabular}
\label{table:plants}
\end{table}

\begin{table}[tbp]
\centering
\caption{Scintillator mixtures used in the different periods of \dsf\ running. Fractions of \pc\ and \tmb\ are mass fractions.}
\begin{tabular}{c | c | c}
 \hline
   Period & Experimental Phase & Scintillator mixture \\
  \hline
  \hline
  Nov. 2013 -- June 2014 & \phaseone~\cite{Agnes:2015gu} & \pc\ 50\%, \tmb\ 50\%, \lsvppoconcorig\ \ppo\ \\
  \hline
  June 2014 -- Sept. 2014 &  & \pc\ $\sim$100\%, \tmb\ $\leq 0.05$\%, $2.5$\,g$/$L \ppo\ \\
  Sept. 2014 -- Dec. 2014 &  & \pc\ $\sim$100\%, \tmb\ $\leq 0.05$\%, $\leq 0.1$\,g$/$L \ppo\ \\
  Dec. 2014 -- Jan. 2015 &  & \pc\ $\sim$100\%, \tmb\ $\leq 0.05$\%, $0.7$\,g$/$L \ppo\ \\
  Jan. 2015 -- Feb. 2015 &  & \pc\ 95\%, \tmb\ 5\%, $0.7$\,g$/$L \ppo\ \\
  \hline
  Feb. 2015 -- present & \phasetwo~\cite{Agnes:2015_uar} & \pc\ 95\%, \tmb\ 5\%, \lsvppoconcnow\ \ppo\ \\
  \hline
 \end{tabular}
\label{table:scint_periods}
\end{table}

The distillation and nitrogen stripping facilities were developed based on the experience gained with the similar facilities in Borexino~\cite{alimonti_liquid_2009, Benziger:2009ga}.
Figure~\ref{fig:plants} is a block diagram of the fluid handling plants. The purification plant is composed of a vacuum distillation column and a packed stripping column, and was used both for the scintillator and for the master solution (described later) purification; the main parameters are reported in table~\ref{table:plants}.

All \dsf\ plants and piping had to match Borexino requirements in term of leak tightness and cleanliness, and Borexino cleaning procedures were also adopted~\cite{lombardi_borexino:_2014}. 
In addition, since \tmb\ is highly hygroscopic, special care was taken to ensure that all surfaces were dry after cleaning.
The \dsf\ plants were connected to the Borexino plants, allowing the use of the Borexino storage area and purification plant in addition to the \dsf\ facilities.

The scintillator mixing took place in two stages. First, a \emph{master solution} was prepared by distilling the total mass of \ppo\ and dissolving it in distilled \pc. Second, \pc\ and \tmb\ were separately distilled and mixed together in-line along with the master solution to ensure that all components were homogeneously combined.

The \lsv\ was reconstituted between \phaseone\ and \phasetwo\, in order to reduce the \cfor\ contamination.
The first step was to remove the high-\cfor\ \tmb, which was accomplished using the Borexino distillation plant.
The Borexino plant was used because the \dsf\ plant only has a single stage and was not effective in the separation of \pc\ and \tmb. 
The reconstitution began in June 2014, with five repeated distillations of the entire 30\,tonnes of \pctmb-\ppo\ mixture. 
The \tmb\ was the light product of the distillation and was extracted at the top of the column and sent to one of the storage vessels for disposal.
The \pc\ and the \ppo\ were extracted from the bottoms and sent back to the detector.
The operation was done in-loop in the detector, and the removed \tmb\ was replaced with pre-purified \pc\ from  the storage area.
The distillation process reduced the \tmb\ by a factor of 1000.

After the \tmb\ removal, the light yield of the scintillator dropped by a factor of 2 due to the presence of heavy contaminants in the bottom of the distillation column, and another scintillator purification became necessary. This purification was performed using the Borexino distillation plant followed by nitrogen stripping.
This time the light product was the \pc, which was sent back to the \lsv. The heavy product was \ppo\ with some \pc, which was recovered as the master solution. 
The \pc\ purification increased the light yield of the \lsv\ from $\sim0.25$\,PE$/$keV to $\sim0.4$\,PE$/$keV.
As a side effect of the distillation, the \ppo\ was removed from the mixture.

The entire scintillator inventory was cycled through external pumps to allow the \ppo\ to be re-added in two aliquots.
This was done using the master solution, which was purified with the \dsf\ purification plant and added in-line with the scintillator.
This re-addition took place in December 2014 and February 2015.
The first \ppo\ addition campaign restored the \ppo\ concentration from a few ppm to 680\,$\pm$\,30\,ppm ($\sim$\lsvppoconcint).
The additional \ppo\ increased the light yield of the \lsv\ from $\sim0.4$\,PE$/$keV to $\sim0.55$\,PE$/$keV with no increase of \cfor\ or other backgrounds.

In January, 2015, further operations were performed to add the low-\cfor\ \tmb. 
For cost reasons, the new \tmb\ was added at a lower mass fraction of 5\%, rather than 50\%.
During the \tmb\ addition campaign, which took place between January 8--15, 2015, a total of $1.44$\,tonnes of \tmb\ were added to the \lsv.

In February, 2015, the second \ppo\ aliquot was added.
This increased the \ppo\ concentration from  \lsvppoconcint\ to \lsvppoconcnow.
The further addition of \ppo\ slightly decreased the light yield of $\beta$ and $\gamma$ interactions 
from $\sim 0.55$\,PE$/$keV to $\sim 0.53\,\pm\,0.03$\,PE$/$keV, but also slightly increased the light yield for $\alpha$ + \lith\ products, by decreasing the effects of $\alpha$ + \lith\ scintillation quenching~\cite{Berlman}.
More details on the light yield measurements and changes can be found in section~\ref{sec:performances}. 

After adding the clean \tmb, the \cfor\ decay rate in the \lsv\ dropped from its original value of \lsvcforratehigh\ to \lsvcforratelow.
Table~\ref{table:scint_periods} summarizes the various scintillator mixtures in the different periods of \dsf\ running.

\subsection{Stainless Steel Sphere}
\label{sec:lsv_sss}

An unsegmented Stainless Steel Sphere (SSS) is both the container of the boron-loaded scintillator and the mechanical support of the \lsvpmtnum\ \lsvpmt\ \lsvpmtsize\ \pmts\ of the \lsv.
The SSS has a diameter of \lsvdiameter, a volume of roughly 30\,m$^3$, and was assembled and welded inside the \wcd\ tank from pre-shaped AISI316L steel plates of 10\,mm thickness.
The SSS is supported by four steel legs 
to keep the sphere's south pole $2.4$\,m above the base of the \wcd\ water tank.
The steel legs are welded to the base plate of the water tank.

The \dsf\ cryostat is inserted through a 1\,m diameter flange at the top of the sphere,
and is suspended by three stainless steel rods with lengths adjustable with sub-millimeter repeatability.
The flange also includes seven feed-throughs to connect, via corrugated steel hoses, all the services for the cryostat (vacuum line, \pmt\ cables, optical fibers, high voltage \tpc\ cables, etc.).

Four straight vertical {6}\,{\inch} tubes connect the inner volume of the sphere to the radon-free clean room on top of the \wcd\ (see figure~\ref{fig:ds50}).
Each tube is welded to the SSS on one end and connects to a port in bottom of the clean room on the other end.
One of these ports is currently in use for the source calibration system (section~\ref{sec:CALIS}), and the others are available for other uses. Three additional {10}\,{\inch} flanges, two on the upper and one on the lower hemisphere, are equipped with CCD cameras for visual monitoring of the detector status during operations and between periods of normal running. 

\subsection{Lumirror reflector}
\label{sec:lumirror}

\begin{figure}[tb]
 \centering
 \includegraphics[width=0.8\linewidth]{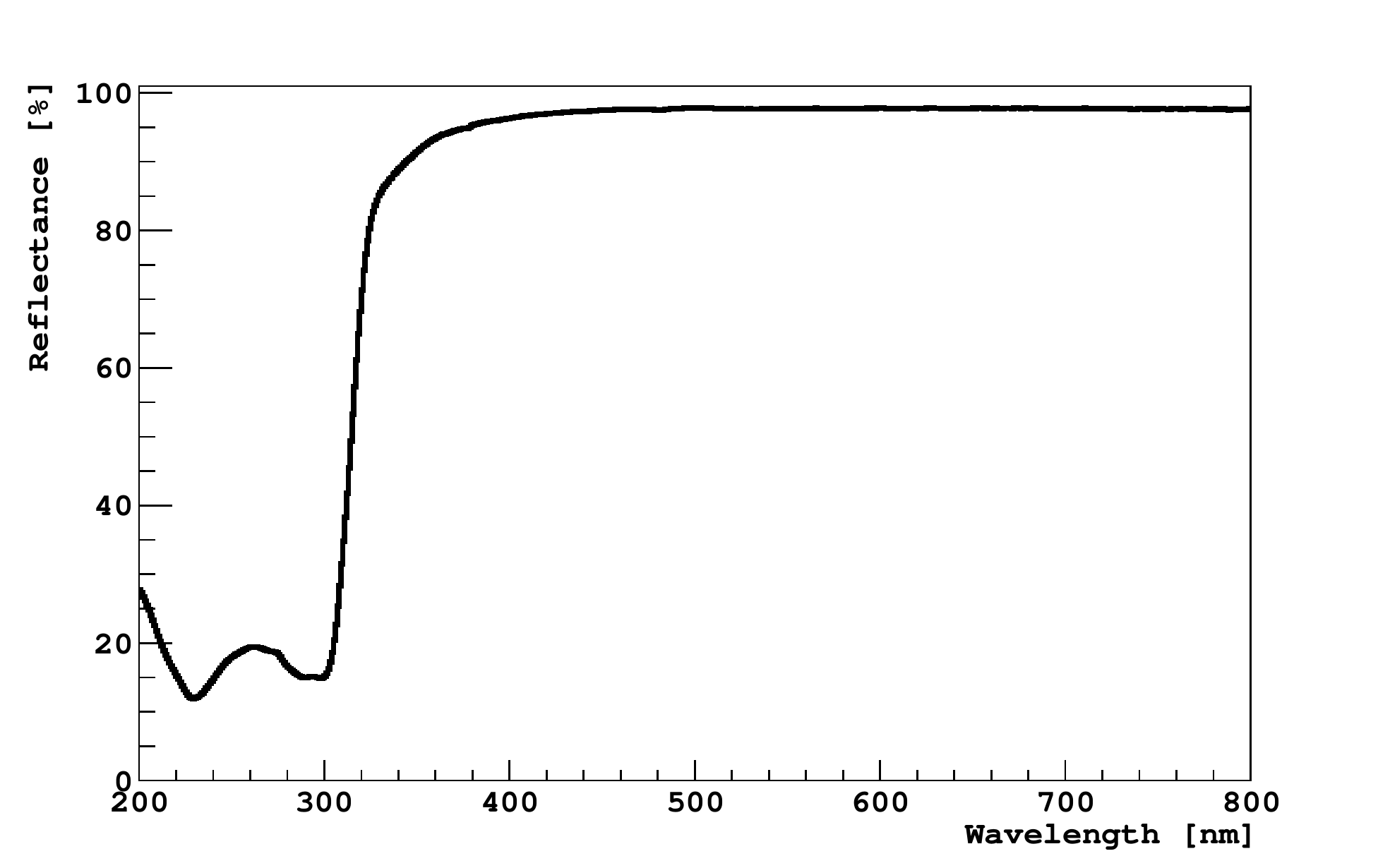}
 \includegraphics[width=0.8\linewidth]{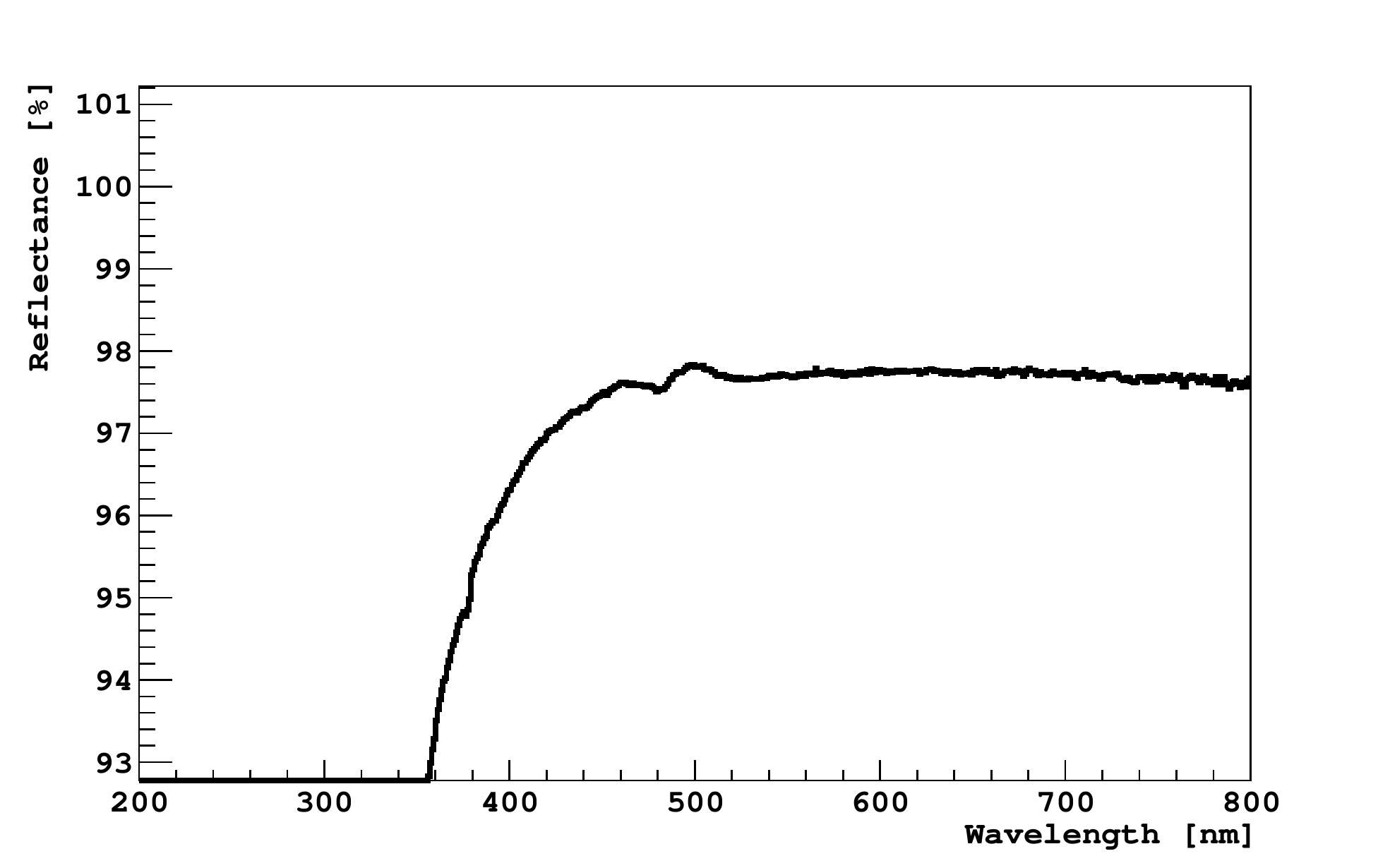}
 \caption{The reflectance measured for Lumirror E6SR using a Perkin Elmer Lambda-650 spectrophotometer. There is a sharp cutoff at wavelengths shorter than $\sim$350\,nm. The bump in reflectance around 260\,nm is due to the specular component of the reflectivity. (Top) The reflectance over the full range measured. (Bottom) The reflectance near the maximum value.}
 \label{fig:lumirror}
\end{figure}

Due to cost restrictions the \lsv\ has $\sim$7\% photo-detector coverage. 
To compensate for this,  the inner walls of the \lsv\ have been lined with a reflector.

Many reflectors were initially ruled out due to chemical incompatibility with \pc\ or \tmb. 
Many other reflectors, like Gore-Tex~\cite{gore1976process}, rely on voids with different indices of refraction than the bulk, and therefore lose reflectance when submerged in liquid scintillator~\cite{westerdale_prototype_2015}. 
After some experimentation~\cite{shields_sabre, westerdale_prototype_2015}, Lumirror 188 E6SR was chosen. Lumirror is a highly reflective void-based reflector between two protective layers that allow the reflector to be submerged in scintillator without significant loss of reflectivity~\cite{janecek_reflectivity_2012,sakaguchi2014white}.

A sample of measurements of the Lumirror reflectance taken with a Perkin Elmer Lambda-650 spectrophotometer are shown in figure~\ref{fig:lumirror}. 
As can be seen in this figure, the Lumirror has a maximum reflectance of $\sim$97.5\% for wavelenghts greater than $\sim$350\,nm, and has a steep cutoff at shorter wavelengths, 
below most of the scintillator's emission spectrum. 
Additional tests have shown that while prolonged periods of soaking the Lumirror in scintillator do not significantly decrease the reflectance of the bulk of the Lumirror, there is a very slowly progressing degradation along the edges of about 1 cm per 9 months. This degradation decreases the peak reflectance of the edges to $\sim$83\%~\cite{westerdale_prototype_2015}. 
However, this effect progresses inwards slowly enough to not be a major problem for a prolonged campaign. 

The Lumirror was installed in the \lsv\ as a series of spherical wedges strips in order to line the inner surface of the sphere. 
As an additional measure against the degradation observed along the edges of the Lumirror, the strips were overlapped by $\sim$5\,cm when installed. 

A picture of the \lsv\ with the \lar\ \tpc\ inserted into it is shown in figure~\ref{fig:lsvpic}.
The exterior surface of the \tpc\ cryostat is not covered with Lumirror.
Since the \tpc\ cryostat's electropolished stainless steel is less reflective than the Lumirror,
its presence in the center of the \lsv\ affects the optics of the \lsv.
Scintillation events that happen near the cryostat are likely to lose some light due to absorption at the stainless steel surface. 
Covering the cryostat with a layer of Lumirror in order to reduce this effect was considered,
 but this option was discarded due to the risk of a neutron capturing between the Lumirror and the cryostat, losing all of its light.

\subsection{\lsv\ photomultiplier tubes}
\label{sec:lsv_pmts}

\begin{figure}[tb]
 \centering
 \includegraphics[width=.45\linewidth]{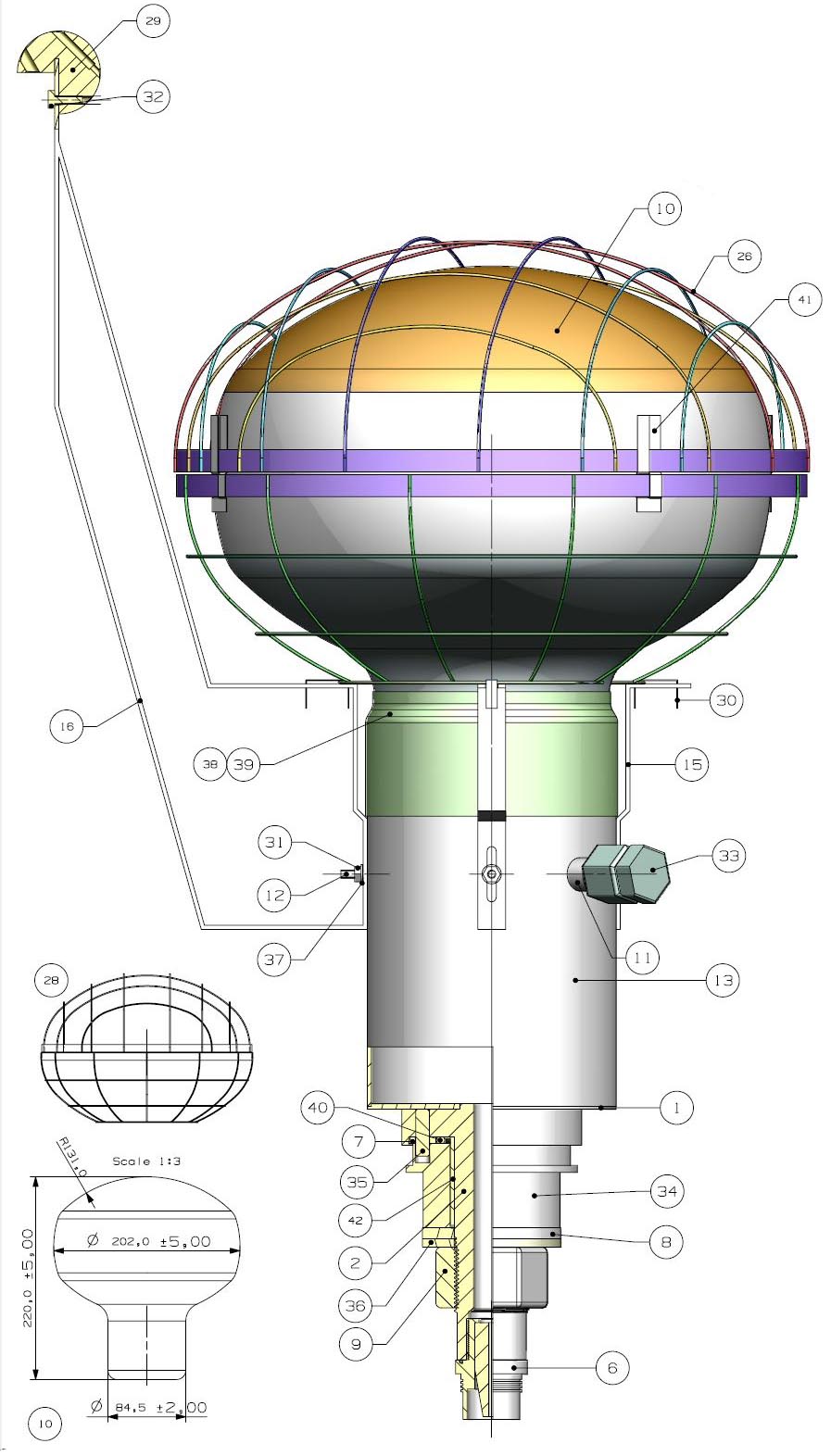}
 \includegraphics[width=.54\linewidth]{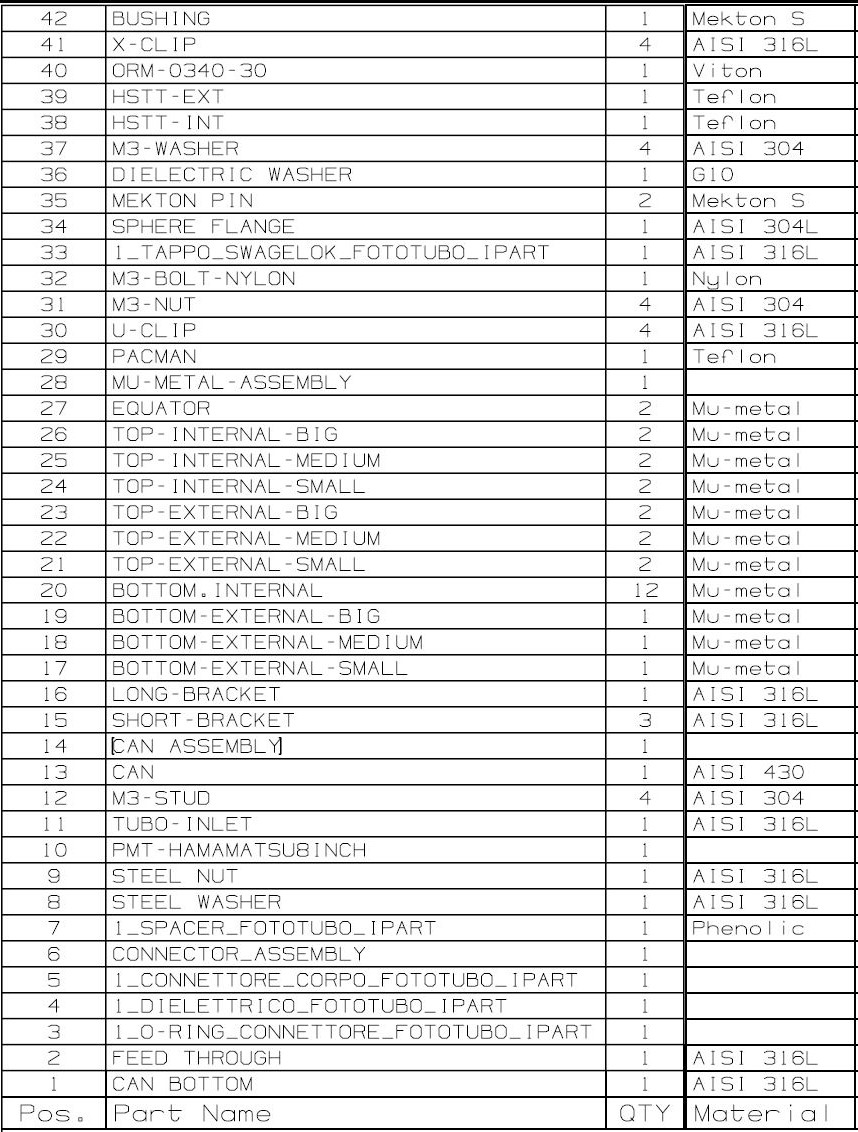}
 \caption{Design of the \lsvpmt\ photomultiplier tube intalled in the \lsv.}
 \label{fig:lsv_pmt}
\end{figure}

\begin{figure}[tb]
 \centering
 \includegraphics[width=.8\linewidth]{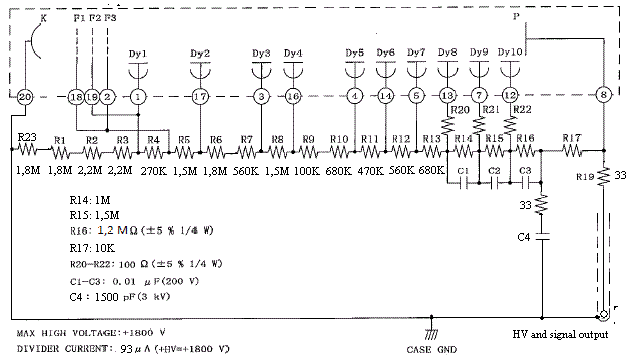}
 \caption{The \lsvpmt\ \pmt\ voltage divider.}
 \label{fig:lsv_pmt_vd}
\end{figure}

\begin{table}[tb]
 \centering
 \caption{Radioactivity of the \lsv\ \lsvpmt\ reported by Hamamatsu. No uncertainties were provided for these measurements. The mass of a single phototube is $\sim$1100 g~\cite{hamamatsu_r5912}.}
 \begin{tabular}{l|cccc}
 \hline
  Isotope             & \pota & \ura-series & \urafive-series & \tho-series \\\hline
  \hline
  Activity [mBq/\pmt] & 649   &  883        & 41              & 110 \\
  \hline
 \end{tabular}
 \label{vpmt_rad}
\end{table}

The \lsv\ is equipped with \lsvpmtnum\, \lsvpmt\ \lsvpmtsize\ \pmts\, with low-radioactivity glass bulbs and high-quantum-efficiency photocatodes (\lsvpmtqe\ average quantum efficiency at \lsvpmtwave)~\cite{hamamatsu_r5912}.
All \lsvpmtnum\ \pmts\ are mounted on the SSS,
as shown in figure~\ref{fig:lsvpic2}. 

The design was based upon the broad experience gained in the  Borexino experiment~\cite{lombardi_optical_2014}.
The \pmts\ were chosen for their low radioactivity glass, low dark-pulse rate, low \aftp\ rate, and a 1\,$\sigma$ transit time spread of $\sim1$\,ns, much smaller than the scintillation light pulse width. The amplitude and timing response performance of single electrons is particularly important due to the need to measure prompt coincidences with the \lar\ \tpc\ with a very low threshold. 
Furthermore, in order to minimize the background from \pmt\ noise, the \pmts\ have to feature a low dark rate and have a low probability of \pmt\ after-pulsing.
All of the \pmts\ were tested in the photomultiplier tube testing facility of the Borexino experiment at LNGS~\cite{brigatti_photomultiplier_2005}.

A diagram of the \lsvpmt\ is shown in figure~\ref{fig:lsv_pmt}. The \lsvpmt\ tube has a hemispherical photocathode with a radius of curvature of about 11\,cm and a diameter of about $20.2$\,cm, resulting in a surface area of 380\,cm$^2$. The 
bialkali photocathode is made of CsKSb. 
The bulb of the tube is made of low radioactivity glass, giving the \pmts\ radioactivity levels shown in table~\ref{vpmt_rad}, as reported by Hamamatsu.
The multiplier structure consists of 10 linear focused dynodes (BeCu), as shown in figure~\ref{fig:lsv_pmt_vd}.

The \pmts\ are installed from inside the sphere, connected by feed-throughs uniformly spaced across the sphere wall to a single cable outside the sphere carrying both signal and high voltage. A large fraction of the cable is submerged in water in the \wcd. The back-end sealing of the \pmts\ has been designed to be compatible with operation in \pc, \tmb, and water.
To assure a reliable operation of the \pmts\ in such a complex environment, it was necessary to study and develop an encapsulation of the neck of the bulb and the divider. 

The base and the neck of the tube are enclosed in cylindrical stainless steel housing with an external diameter of 90\,mm.
This housing is fixed to the glass of the tube neck using the \pc-compatible EP45HT epoxy resin from Master Bond, which acts as a structural adhesive as well as a protective barrier against \pc\ ingress.
The end-cap of the cylindrical housing was welded and helium leak tested. It forms the feed-through which is inserted into a hole on the sphere surface and is secured with a rear nut.
In this way, the front part of the \pmt\ assembly is immersed in scintillator and the rear part in water.
Phenolic resin is used as an insulating material that electrically decouples the device to avoid ground loops and also acts as a groove for the Viton O-ring, assuring the tightness between the \pmt\ feed-through and the sphere.

All \pmt\ feed-throughs have been tested with a helium leak detector.
The feed-throughs were  designed to accommodate the underwater jam-nut connector from the company Framatome~\cite{framatome}. 
The connector is screwed into the feed-through until its O-ring is properly compressed; a further potting with the same epoxy resin mentioned before is also made in order to have a second barrier against water infiltration.
The space inside the cylinder is filled with an inert organic oil via a 10\,mm pipe port sealed afterwards with a Swagelok cap.
This mineral oil prevents water from condensing on the divider without stressing the very delicate joints between the metal pins and the glass.

The last barrier against \pc\ is provided by a heat shrink Teflon tube glued with epoxy resin between the glass neck of the \pmt\ and the steel can.
In order to fully benefit from this barrier, we exploited a patented technology of the Gore company (Tetra-Etch, a trademark of W.L. Gore \& Associates~\cite{gore_tetra_etch}) which employs a preliminary surface etching of the Teflon film and allows a strong Teflon adhesion on other surfaces.

The \lsvpmt\ \pmts, due to their large size, are sensitive to the Earth's magnetic field. Thus, it is necessary to shield them with a custom-made $\mu$-metal grid ($1.0$\,mm thick wires) placed around the photocathode. Accelerated ageing tests showed that this material, in contact with \pc, strongly catalyzed the scintillator oxidation.
It is therefore necessary to cover the $\mu$-metal with a 20\,$\mu$m thick lining of clear phenolic paint from Morton~\cite{sterilkote}.
The support of the $\mu$-metal grid is sketched in figure~\ref{fig:lsv_pmt}:
it is mounted with four holders at 90 degrees, connected to the housing of the tube with  screws welded onto it.

In order to determine the energy of scintillation events, it is important to know the total charge collected by each \pmt.
For this reason, charge calibration of the \pmts\ is required.
The light emitted by an external picosecond 405\,nm laser source is carried simultaneously to each \pmt\ by a dedicated system of optical fibers via 10 custom-designed optical feed-throughs, which are further split into \lsvpmtnum\ individual fibers each coupled to a \pmt. The design has been based on the experience gained in the development of the Borexino experiment~\cite{caccianiga_multiplexed_2003}.

The system's cables and connectors will be completely immersed in high purity water for many years,
so we selected designs and materials  developed for submarine applications, following the experience gained in Borexino~\cite{2008apsp.conf..214L}.
The fundamental requirements are material compatibility and electrical performance. 
Plastics must have a very low water absorption coefficient to withstand long exposure to high purity water.
The entire design must include multiple barriers against radial and axial diffusion of water along or through the cable jacket.
Materials for the sealing, the cables, and the connectors as well as the other mechanical parts require the evaluation of possible effects caused by the concurrent presence of different materials together, such as the development of galvanic couples and localized corrosion, as well as the release of impurities into the water.
With the \pmts\ in the scintillators, the radioactivity level of the materials is also a major concern.

A single-cable transmission line is used to conduct both the signal and the high voltage (HV), connected to a 50\,$\Omega$ front-end where the decoupling of the HV and the signal is accomplished. 
The cables are custom made 50\,$\Omega$ RG~213 coaxial  cables.
The outer jacket is made of solid extruded high density polyethylene with a laminated copper foil, bounded to the braid with a copolymer coating serving as a second barrier.
All the cables have an electrical length of $246.8$\,$\pm$\,$0.25$\,ns  ($\sim$49\,m), and some of this length is under water.
Framatome  provided  stainless steel connectors suitable for RG~213 cables~\cite{framatome} together with the mounting structure of the tube housing. 
Cable impedances are 50\,$\pm$\,2\,$\Omega$. 
All of the O-rings are made of Viton, which has been found to interact with neither \pc\ nor \tmb.

The HV is supplied to the \pmts\ by CAEN A1536 HV boards~\cite{caen_a1536}, housed in a CAEN Mainframe SY4527~\cite{caen_sy4527}.
Typical operating high voltages are between 1600 and 1800\,V.

The schematic diagram of the \pmt\ voltage divider is shown in figure~\ref{fig:lsv_pmt_vd}.
The signal at the \pmt\ output due to a single photo-electron (SPE) is a pulse of about 12\,mV on a 50\,$\Omega$ load, with FWHM of $\sim$10\,ns. Detailed information on the front-end and digital electronics used to acquire the \pmt\ pulses can be found in~\cite{Agnes:2015dsvetoelec}.

All but two of the \lsvpmtnum\ \pmts\ of the \lsv\ are, at the time of writing, operating correctly.
The two failures in the \pmts\ of the \lsv\ occurred while filling the \wcd, when the water level reached their submarine cable connectors.

\subsection{Calibration system}
\label{sec:CALIS}

\begin{figure}[tbp]
\centering
\includegraphics[width=0.50\textwidth]{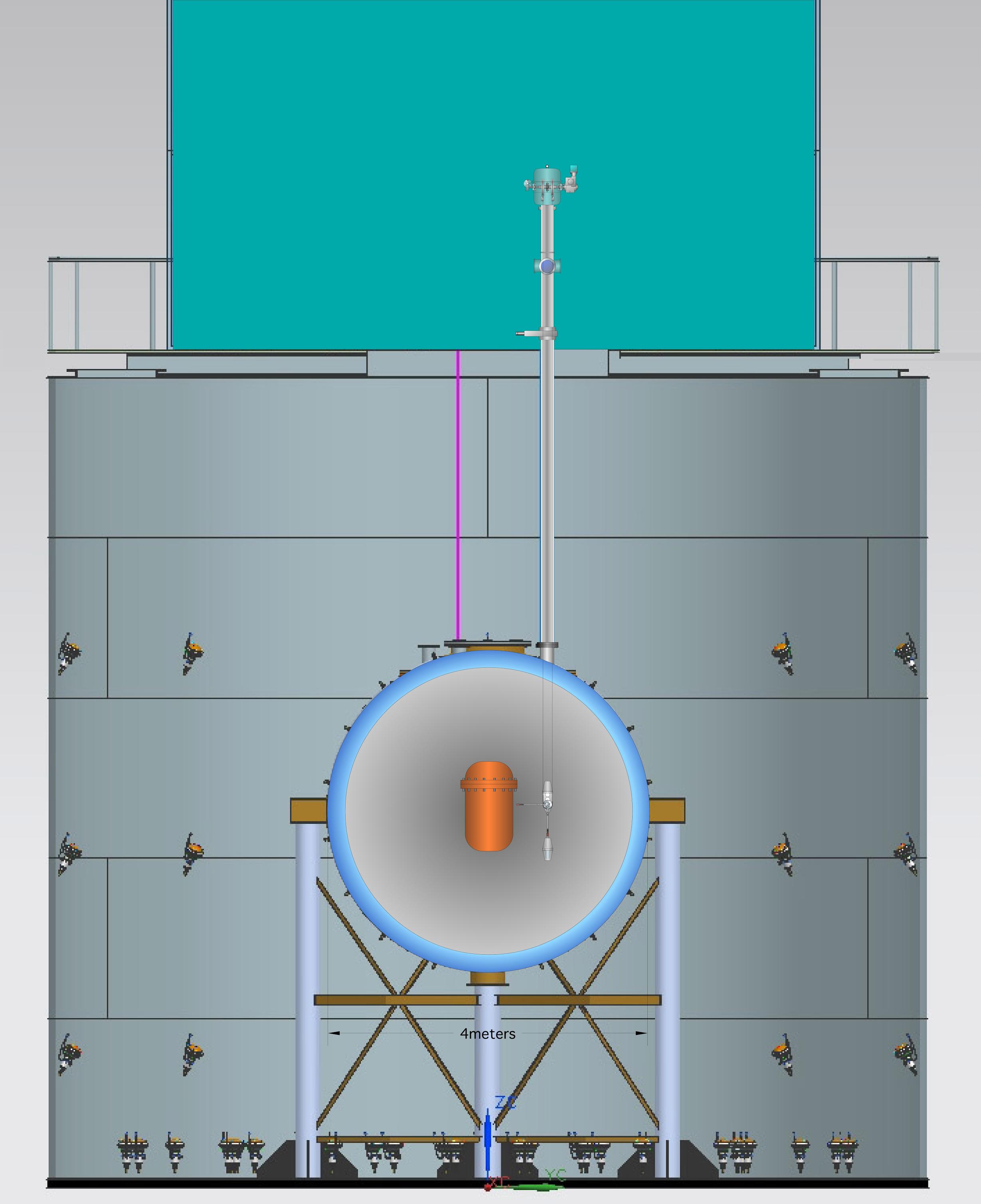}
\includegraphics[width=0.44\textwidth]{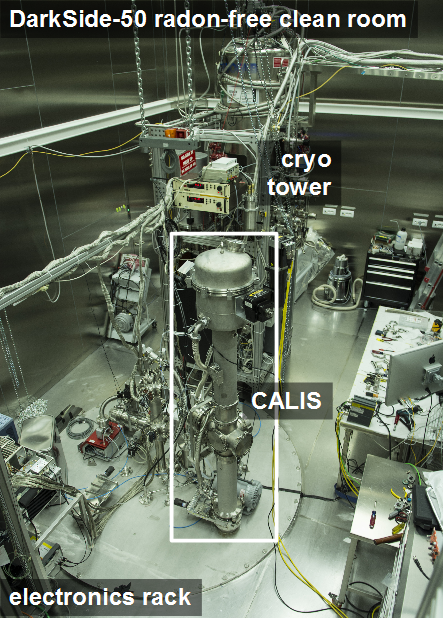}
\caption{\textit{Left}: Sketch of CALIS installed inside the radon-free clean room CRH atop the \wcd\ and with the deployment device next to the cryostat. The source arm is articulated.
\textit{Right}: Photograph of CRH after installation of CALIS.
\label{fig:CALIS}}
\end{figure}

\begin{figure}[tbp]
\centering
\includegraphics[width=0.51\textwidth]{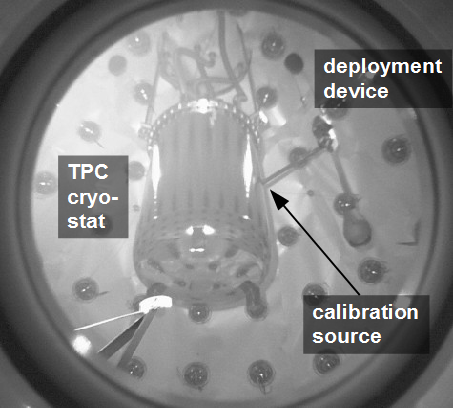}
\includegraphics[width=0.47\textwidth]{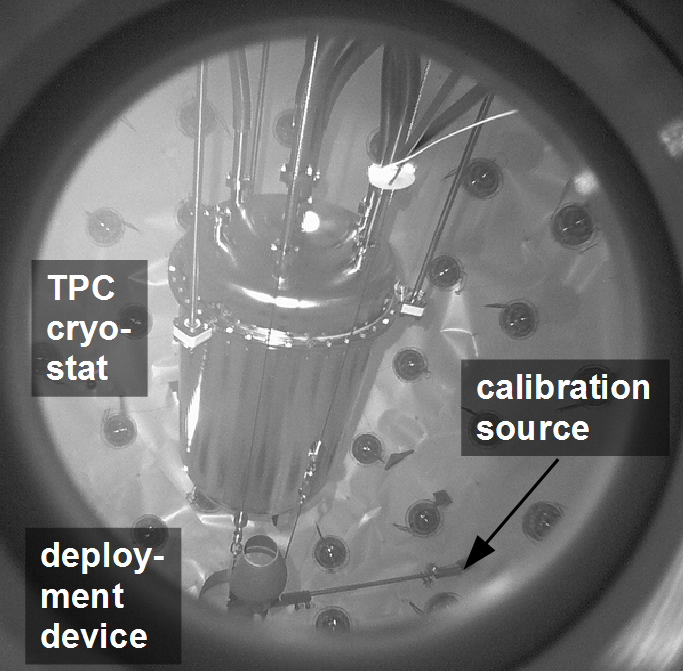}
\caption{Photographs taken with cameras looking into the \lsv. \textit{Left:} CALIS with the source in contact with the cryostat of the \lar\ \tpc. \textit{Right:} CALIS with the \AmBe\ source rotated away from the \lar\ \tpc\ in order to study position dependent effects. 
\label{fig:CALIS:Pig}}
\end{figure}

A CALibration Insertion System (CALIS) is used to deploy radioactive sources to calibrate both the \lar\ \tpc\ and the \lsv.
CALIS was constructed, tested at Fermilab and LNGS, precision cleaned, and installed in  \dsf\ during summer, 2014 (figure~\ref{fig:CALIS}).
Since then it has been used for extended calibration campaigns of both the \tpc\ and \lsv\ involving radioactive \gr\ (\cobaseven, \cesium, \barium) and neutron  (\AmBe) sources, with further campaigns planned.
A detailed description of CALIS hardware will be the subject of a future paper; in this document, we will focus primarily on aspects that influence the \lsv\ calibration.

The source holder housing the radioactive source is at the tip of an arm, which hangs vertically downwards during the insertion of the deployment device into the \lsv\ and is articulated to a horizontal position when its goal height is reached.
The arm including the source holder has a total length of 62\,cm, which allows the source holder to make physical contact with the \lar\ \tpc\ cryostat, thereby bringing the radioactive source as close as possible to the \lar\ \tpc\ active volume (figure~\ref{fig:CALIS:Pig}, left).
This feature is particularly useful to calibrate the \lsv\ response to neutrons that are captured in the \lsv\ volume close to the \lar\ \tpc\ cryostat.
The arm can been rotated away in order to study position-dependent effects (figure~\ref{fig:CALIS:Pig}, right),
which are expected due to the absorption of light on the \lar\ \tpc\ cryostat.

\section{The Water Cherenkov Veto}
\label{sec:wcd}
\begin{figure}[tb]
 \centering
 \includegraphics[width=0.8\linewidth]{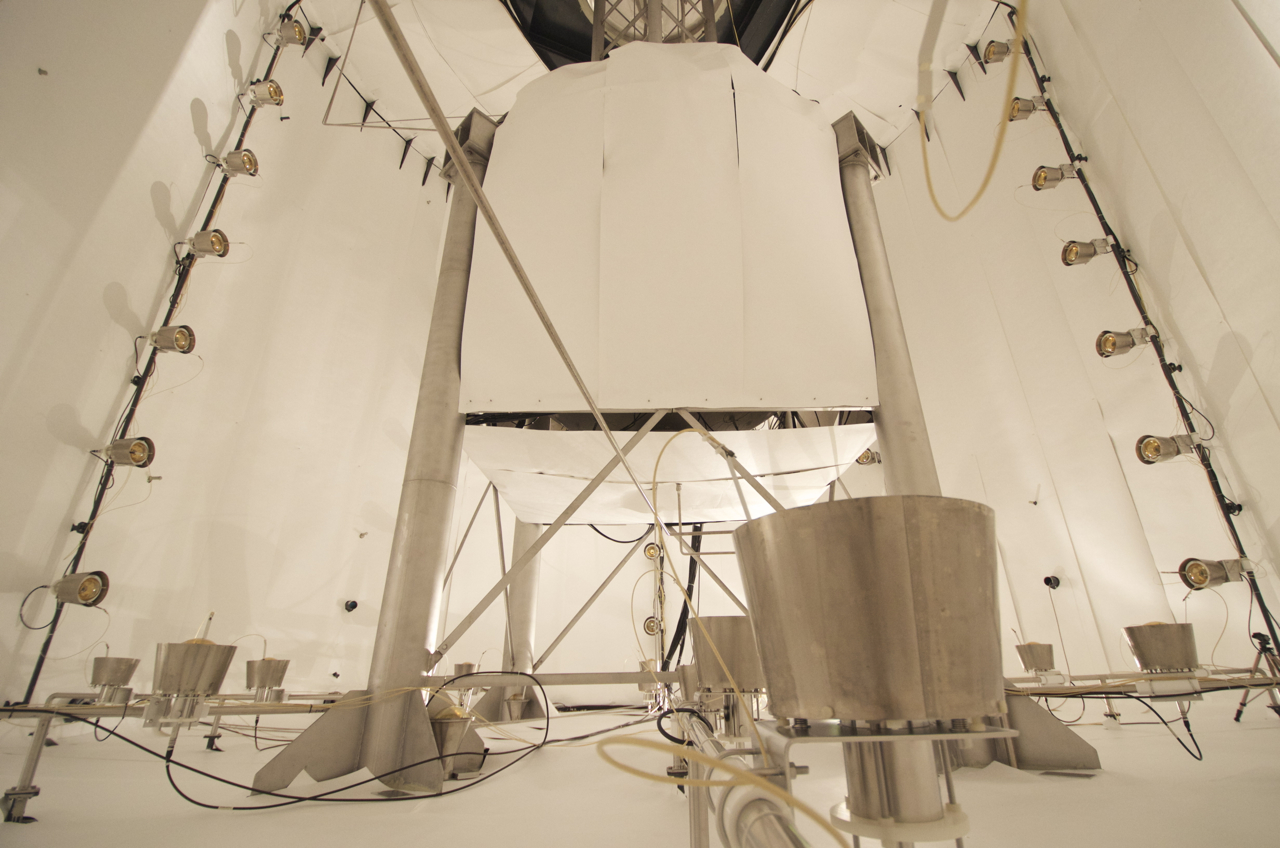}
 \caption{The inner surface of the Water Cherenkov Veto covered with layers of Tyvek.
 The Liquid Scintillator Veto, installed on four stilts and also covered in Tyvek, can be seen in the middle of the water tank.}
 \label{fig:wtpic}
\end{figure}

The Water Cherenkov Veto (\wcd) is a powerful shield against external background (\gr s and neutrons from the rock), and is also used as a Cherenkov muon detector. The muon flux
at the \lngsdepth\ depth of the LNGS, although reduced
from that at the Earth's surface
 by a factor $\sim10^6$, is of the order of \lngsmuflux~\cite{bellini_cosmic-muon_2012}.
This corresponds to about 2000 muons per day crossing the \wcd, about 380 muons per day crossing the \lsv, and about 4 muons per day crossing the \lar\ \tpc.
Cosmogenic muons can produce high energy neutrons~\cite{Bellini:2013kr, empl_fluka_2014}, which can penetrate several meters of shielding. 
In order to avoid backgrounds from these high energy neutrons, the \wcd\ acts as a veto to detect the muons that may produce them and therefore leave a detectable coincident signal. 
The \wcd\ is  equipped with \ctfpmtnum\  \ctfpmtsize\ \pmts: 56 in 8 columns on the cylindrical tank wall and 24 on the floor.
These \pmts\ collect the Cherenkov light emitted by muons or muons' electromagnetic shower products in the water.
The \wcd\ uses the water tank from the Borexino Counting Test Facility (\ctf)~\cite{alimonti_large-scale_1998}, and the design draws on the muon detector of the Borexino experiment. The water used to fill the \wcd\ was purified using the Borexino water purification plant, described in~\cite{giammarchi_water_2014}, which purifies water to the levels shown in table~\ref{table:water}.
The process of filling the \wcd\ with water began on October 2 and ended November 15, 2013.

\begin{table}
\centering
\caption{Water contamination levels before and after purification with the Borexino water purification plant, as reported by~\cite{giammarchi_water_2014}}
\begin{tabular}{l|cc}\hline
Isotope & LNGS Water [Bq/kg] & Purified Water [Bq/kg] \\\hline\hline
$^{238}$U & $1\times10^{-3}$ & $3\times10^{-7}$\\
$^{226}$Ra & 0.3 & $<1\times10^{-6}$\\
$^{222}$Rn & 10 & $<3.4\times10^{-6}$\\
$^{232}$Th & $1\times10^{-3}$ & $3\times10^{-7}$\\
$^{40}$K & $1\times10^{-3}$ & $<1.6\times10^{-6}$\\\hline
\end{tabular}
\label{table:water}
\end{table}

Figure~\ref{fig:wtpic} shows a photograph of the inside of the \wcd.
The water tank of the \wcd\ is a cylinder with a diameter of \ctfdiameter\ and a height of \ctfheight\ for a total volume of $\sim1000$~m$^3$.
It is built from carbon steel and internally protected from corrosion by a layer of Permatex resin.
In order to improve the light collection efficiency, the interior surface of the \wcd\ and the exterior of the \lsv\ sphere are covered with a layer of Tyvek~\cite{dupont_tyvek1082d, janecek_reflectivity_2012}.

Since the background signals in the \wcd\ are very small compared to the large Cherenkov signal produced by the muons, light collection efficiency is not as crucial as it is in the \lsv, allowing us to reuse the \pmts\ from \ctf\ and use cheaper reflector without sacrificing detector performance. 

\subsection{Tyvek reflector}
\label{sec:wtrefl}


The inner surfaces of the \wcd, including the outer surfaces of the \lsv, are covered with a layer of a  Tyvek-based reflector supplied by the Daya Bay collaboration~\cite{daya_bay_collaboration_muon_2015}. 
This reflector consists of two layers of DuPont Tyvek 1082D bonded together by a layer of polyethylene.
The reflectivity of this material has been measured to be greater than 96\% in air and 99\% in water, for light of 300--800 nm in wavelength~\cite{daya_bay_collaboration_muon_2015}.

\subsection{\wcd\ photomultiplier tubes}
\label{sec:wt_pmts}
The \wcd\ is lined with \ctfpmtnum\ 20\,cm diameter  \ctfpmt\ \ctfpmtsize\ \pmts. 
These are the same \pmts\ used by the \ctf\ experiment~\cite{alimonti_large-scale_1998}, after 15 years of continuous operation. 
These \pmts\ have a peak quantum efficiency of $\sim$25\% at 380\,nm and a dark rate of $\sim$2500\,Hz.
All but five of the \pmts\ in the \wcd\ are, at present, operating correctly.

Similar to the \lsv\ \pmts\, the \wcd\ \pmts\ are each equipped with an optical fiber pointing at the photocathodes, and are connected to the same laser system as the \lsv\ \pmts.
Each \pmt\ is surrounded by a conical light collector, made of UV-transparent acrylic coated with layers of silver and copper with a protective layer of acrylic painted on top~\cite{alimonti_large-scale_1998}. These light collectors increase the effective coverage of the \pmts.

\section{Event reconstruction}
\label{sec:reconstruction}

The analog signals from the \lsv\ and \wcd\ \pmts\ are treated identically. The analog signal is decoupled from the HV, and it is amplified by a factor ten by a custom made front-end module.
The amplified signals are sampled by commercial National Instruments \pxie\ 5162 digitizers (10 bit, $1.25$ GHz~\cite{ni_pxie5162}).
Detailed information on the electronics and data acquisition of the veto system of \dsf\ can be found in~\cite{Agnes:2015dsvetoelec}.

During the WIMP search data taking, the trigger of the veto is initiated by the \lar\ \tpc.
When a trigger is received, the veto digital electronics record data in an acquisition window with a width of several neutron capture times in the \lsv.
During \phaseone, when the neutrons had a capture time of $\sim2.2\,\mu$s, the data acquisition window was usually $70\,\mu$s. In this phase, the length of the acquisition window was made much longer than necessary  in order to be able to detect neutrons that captured in \tpc\ materials, such as $^{19}$F or $^{56}$Fe. During \phasetwo, when the neutrons have a capture time of $\sim22\,\mu$s in the scintillator, the data acquisition window was first set to $140\,\mu$s then extended to $200\,\mu$s.

As an alternative to the \lar\ \tpc\ trigger, it is possible to use the \lsv\ itself as a trigger input.
We refer to this configuration as the \lsv\ \emph{self-trigger}. 
The \lsv\ self-trigger configuration has been used in some special runs, like in some of the \AmBe\ source acquisitions (see section~\ref{sec:reco-ambe}) or runs for \radon\ activity evaluation (see section~\ref{sec:bipo}).
The self-trigger is a majority trigger which requires a configurable number of \pmts\ of the \lsv\ to exceed a threshold in amplitude equivalent to $\sim$0.5\,PE within a configurable time window.

At the data acquisition level, we store \pmt\ pulses within the data acquisition window using a zero-suppression algorithm which only records waveform samples which exceeded an amplitude threshold plus a programmable number of samples before and after the threshold crossings.
During WIMP search data taking, the zero-suppression threshold is usually set to a level about one fourth the amplitude of a single-photoelectron pulse; 20 samples (corresponding to 16\,ns) are recorded before and after the pulse.
If the threshold is crossed again before the specified number of post-samples have been recorded, the algorithm waits for the signal to go below threshold and starts counting down from the final threshold-crossing.

The \dsf\ event reconstruction software is built using the Fermi National Accelerator Laboratory's \emph{art} framework~\cite{green_art_2012}.
The software reconstructs the \pmt\ signals from the raw data for every trigger then builds the separate sum waveforms for the \lsv\ and \wcd\ detectors.
Waveform integrals in regions of interest are computed to evaluate the energy deposited in the scintillation (or Cherenkov) events.

\subsection{Pulse reconstruction}\label{sec:pulse_recon}
Pulses are defined as the segments of each waveform for each channel that survive zero-suppression.
Offline software procedures correct the pulses for effects of the electronics,
such as residual DC offset that may affect the waveform pedestal, potential saturation of the  pulse height due to the finite vertical range of the digitizer, and differences in the \pmt\ gains on different channels.

To account for the residual DC offset of each veto channel, a baseline is determined and subtracted from the pulse waveform.
The first 15 samples of each pulse are averaged to define a baseline, which is then subtracted from the waveform of that pulse.

The maximum value of the amplitude range of the waveform digitizer is approximately equal to the amplitude of 7 overlapping SPE pulses.
If many photoelectrons pile-up on top of each other, their total amplitude may exceed the amplitude range of the waveform
digitizer. This would result in the pulse saturating at the maximum amplitude.
An event like this may, for example, be due to a scintillation event in the \lsv\ close to a \pmt.
The offline reconstruction software classifies the pulse as saturated if it finds
a minimum of 3 consecutive samples all having the same amplitude value in the upper
90\% of the amplitude range.
If a pulse is saturated, the offline reconstruction software uses a triangle to approximate the true shape of the pulse peak. 
Two vertices of the triangle are the first and last saturated samples.
The third vertex is the intersection of two lines that extrapolate the rising and falling edge of the pulse.
The area of the triangle is added to the integral of the pulse waveform.
A triangle approximation is used instead of a more sophisticated one (like a Gaussian, or an interpolated \pmt\ pulse template) in order to increase the processing  speed.
The triangular approximation has been tested by artificially saturating non-saturated pulses in the range between 4 and 7\,PE in the reconstruction software.
The integral of the pulse after the saturation correction and the integral before the artificial saturation display a linear trend, with the intercept compatible with zero and correlation coefficient higher than 99\%.

\subsection{Laser calibration}
\label{sec:laser_calib}
To account for different \pmt\ gains, the waveforms of each channel are scaled by the corresponding
SPE mean of that channel.
As described in section~\ref{sec:lsv_pmts}, each \pmt\ in the \lsv\ and \wcd\ has an optical fiber pointing at the photocathode. These optical fibers are attached to a laser which is pulsed at a rate of 500\,Hz. Each digitizer channel is triggered simultaneously by the pulser that drives the laser, opening a 2\,$\mu$s acquisition window for each trigger. The intensity of the laser is set low enough to produce a predominantly single photoelectron distribution on each channel, with a typical occupancy of 1--2\,PE for every 100 triggers on each channel.

Data from these laser runs are processed by selecting pulses that start within a 150\,ns window around the trigger time on each channel to reduce noise from uncorrelated scintillation events, and  then subtracting the baseline of each of these pulses. Each pulse is then integrated, and a histogram of the pulse integrals on each channel is constructed. The SPE mean and relative variance, typically about 10$^{-9}$ V$\cdot$s and $\sim$10\% respectively, are then determined by locating the peak of the SPE distribution and fitting a Gaussian to the peak.

\subsection{Charge integral estimators}
The waveforms of the pulses of each channel, after reconstruction and gain calibration,
are added together to form a sum waveform.
Sum waveforms are produced individually for the \lsv\ and the \wcd.

A scintillation event in the \lsv\ appears as a collection of \pmt\ pulses occurring within a time window of a few hundred nanoseconds.
A similar timescale holds for the Cherenkov events in the \wcd.
A software procedure is necessary to identify both the beginning of the scintillation (or Cherenkov) signal and its end.

We define a \emph{cluster} as the collection of \pmt\ pulses belonging to the same scintillation (or Cherenkov) event
and we call \emph{clustering} the offline software procedure that is applied to identify the cluster (section~\ref{sec:clustering}).
Temporal regions of interest are also defined in the reconstruction software to efficiently identify any signal in coincidence
with the \lar\ \tpc\ scintillation (section~\ref{sec:roi}), while sliding windows are defined to search for regions of maximal charge, in order to identify delayed coincidences (section~\ref{sec:slider}).

\subsubsection{Clustering}
\label{sec:clustering}

\begin{figure}[tb]
 \centering
 \includegraphics[width=0.8\linewidth]{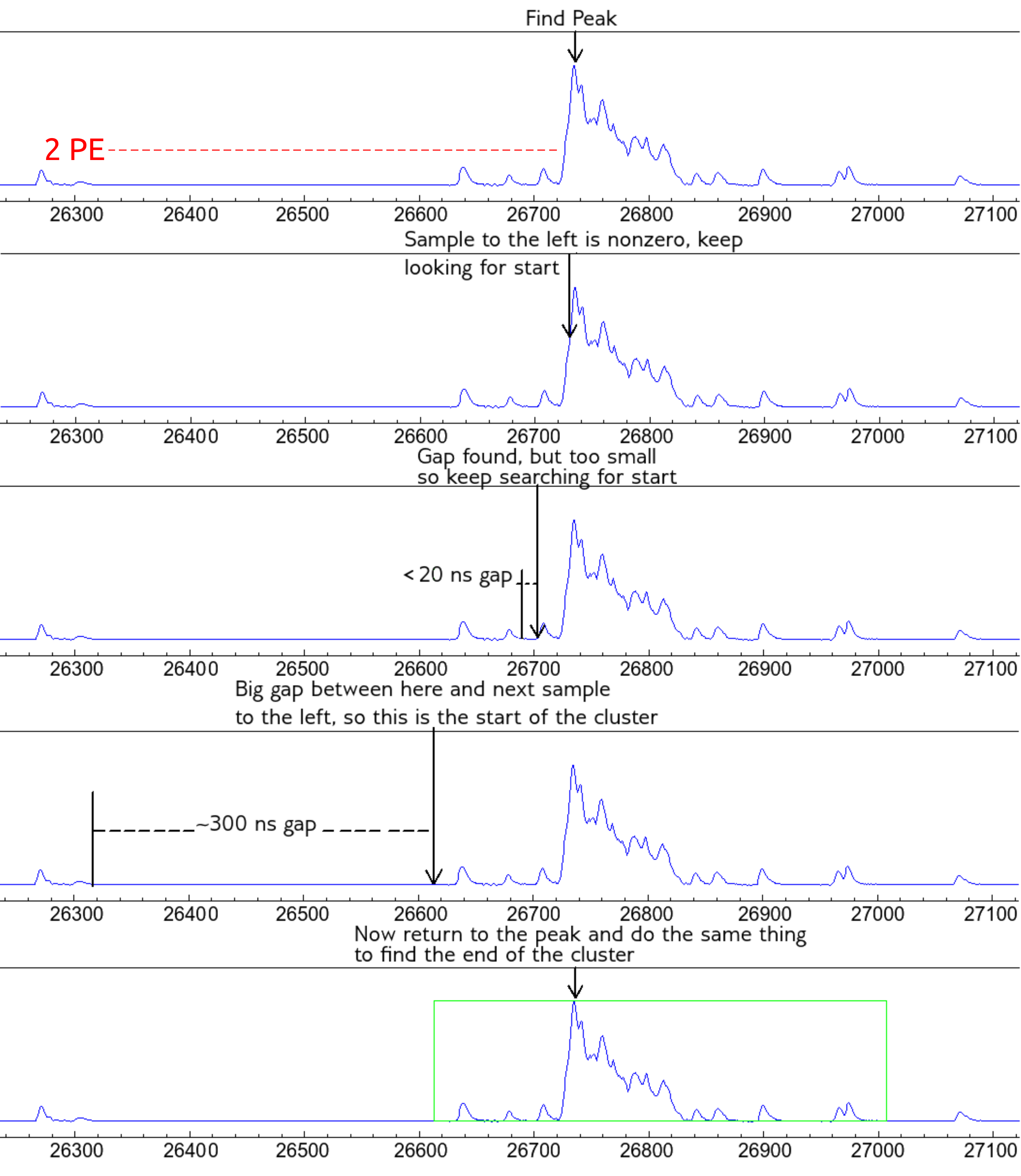}
 \caption{An example of a waveform in the \lsv\ with its corresponding cluster, illustrating the clustering algorithm used. The variable in the horizontal axis is the time in ns with respect to the trigger.}
 \label{fig:cluster}
\end{figure}

A clustering algorithm is applied to the sum waveform to identify scintillation (Cherenkov) events in the \lsv\ (\wcd). 
The clustering algorithm is a top down iterative process that searches for clusters from the largest to the smallest, above some specified threshold (about 2 times the amplitude of a single PE).
The algorithm starts by identifying the highest peak of the sum waveform. 
The algorithm then searches for the start and end time of the cluster by scanning the sum waveform before and after the peak time to identify a gap between pulses greater than 20\,ns.
Once the start and end times of the cluster have been identified, the algorithm identifies the next highest point in the sum waveform that has not yet been assigned to a cluster and repeats.
An example illustrating this algorithm is shown in figure~\ref{fig:cluster}.
This process continues until the highest identifiable peak is below a specified threshold of approximately 2\,PE in amplitude.	

The integral of the sum waveform between the start and end time of a cluster allows the determination of the total energy
deposited in the \lsv\ in the corresponding scintillation event.
In the \wcd, it is used to estimate the amount of Cherenkov light emitted in the event.

\subsubsection{Temporal regions of interest}
\label{sec:roi}
In order to avoid the risk of missing a neutron event in coincidence with the \lar\ \tpc\ due to subtleties in the clustering algorithm (for example, a cluster from a high energy random background shortly after the coincidence signal may pick up the tail of the coincidence signal, lowering the reconstructed charge of the coincident cluster), a series of temporal \emph{regions of interest}, relative to the time the \tpc\ detects a physical event, are used in the dark matter search.
In each region of interest, the \lsv\ sum waveform is integrated between a specified start and end time, for every trigger, regardless of the identification of clusters.
By doing so, the total amount of scintillation light in a specific time window can be calculated reliably.

In the dark matter search, the \emph{prompt} region of interest is especially important. 
This is the time window in coincidence with the \lar\ \tpc\ primary scintillation.
During \phaseone\ of \dsf~\cite{Agnes:2015gu}, the high background rate in the \lsv\ precluded low thresholds and extended time windows.  A 250\,ns wide window was defined around the time of prompt coincidence to detect a neutron thermalization signal greater than 10\,PE. The high background rate caused this cut to accidentally veto $\sim$10\% of all events due to random backgrounds, rather than actual prompt coincidences. If this threshold were placed lower, the accidental veto rate would have become too large to be sustainable. Since the background rate in the \lsv\ in \phasetwo\ is  much lower, the prompt region of interest was extended to a 300\,ns wide window, and the threshold can be placed as low as 1 or 2\,PE while still having an accidental veto rate $\leq 1\%$~\cite{Agnes:2015_uar}.

\subsubsection{Sliding windows}
\label{sec:slider}
A third type of integral is calculated by a \emph{sliding window} algorithm. This algorithm takes a start time, an end time, and a sub-window width (in nanoseconds) and scans the sum waveform between the specified start and end times to find the sub-window of specified width with the greatest integral. This algorithm identifies the largest possible scintillation event that could have happened in the \lsv\ between the specified times for each trigger, without the risk of missing a cluster due to the clustering algorithm splitting a cluster in two or otherwise underestimating a cluster's charge. 
Sliding windows are used in the dark matter search for reliably identifying delayed coincidences with the \lar\ \tpc, such as those expected from a neutron capture.

During \phaseone\ of \dsf~\cite{Agnes:2015gu}, the high background rate completely overwhelmed the low energy region where we expect to see the neutron captures on \borten\ that go straight to the ground state and are not accompanied with a 478\,keV \gr\ (see section~\ref{sec:scint}).
Instead, the goal was to maximize sensitivity to delayed coincidences without having too high of an accidental veto rate.
In order to accomplish this, the \lsv\ acquisition window after the prompt time was divided into two regions: an 8.8\,$\mu$s (4 times the thermal neutron capture in the scintillator mixture of \phaseone) window starting at the prompt \lar\ \tpc\ scintillation time and another going from the end of the first window to the end of the acquisition window (usually 70\,$\mu$s). A slider width of 300\,ns was used during this phase. A veto threshold of 80\,PE in the first window, where neutrons are more likely to capture, and a higher threshold of 110\,PE in the later sliding window, led to a total acceptance loss of $\sim$10\%.

Since in \phasetwo\ the \lsv\ has a much lower background rate, a single sliding window can be defined extending from the prompt time to the end of the acquisition window, usually set to either 140 or 200\,$\mu$s, with a threshold of 3 or 6\,PE, respectively, leading to a total acceptance loss of $\sim$16\%~\cite{Agnes:2015_uar}. The lower threshold significantly increases the detection of the $\alpha$ and \lith\ signals from the neutron capture on \borten, greatly improving the neutron detection efficiency. An additional sliding window extending from the start of the acquisition window to the prompt time was also defined to look for any signal that might precede a neutron scatter in the \lar\ \tpc, for example an external neutron entering the \lsv\ from the \wcd. A threshold of 2\,PE was required for this cut, leading to an acceptance loss of $\sim$0.1\%. During this phase, the lower background rate allowed us to extend the slider width to 500\,ns, ensuring that all scintillation light is collected within the slider.

\section{Veto system performance}
\label{sec:performances}

Once data is reconstructed as described in section~\ref{sec:reconstruction}, it is possible to perform higher level analysis on the reconstructed data to better understand the detectors. 
Most of the analysis discussed here is performed using clusters, as described in section~\ref{sec:clustering}, which can give us information about all of the scintillation events that happened within the \lsv\ acquisition window for each trigger.

\subsection{\lsv\ energy spectrum and backgrounds}
\begin{figure}[tb]
 \centering
 \includegraphics[width=0.8\linewidth]{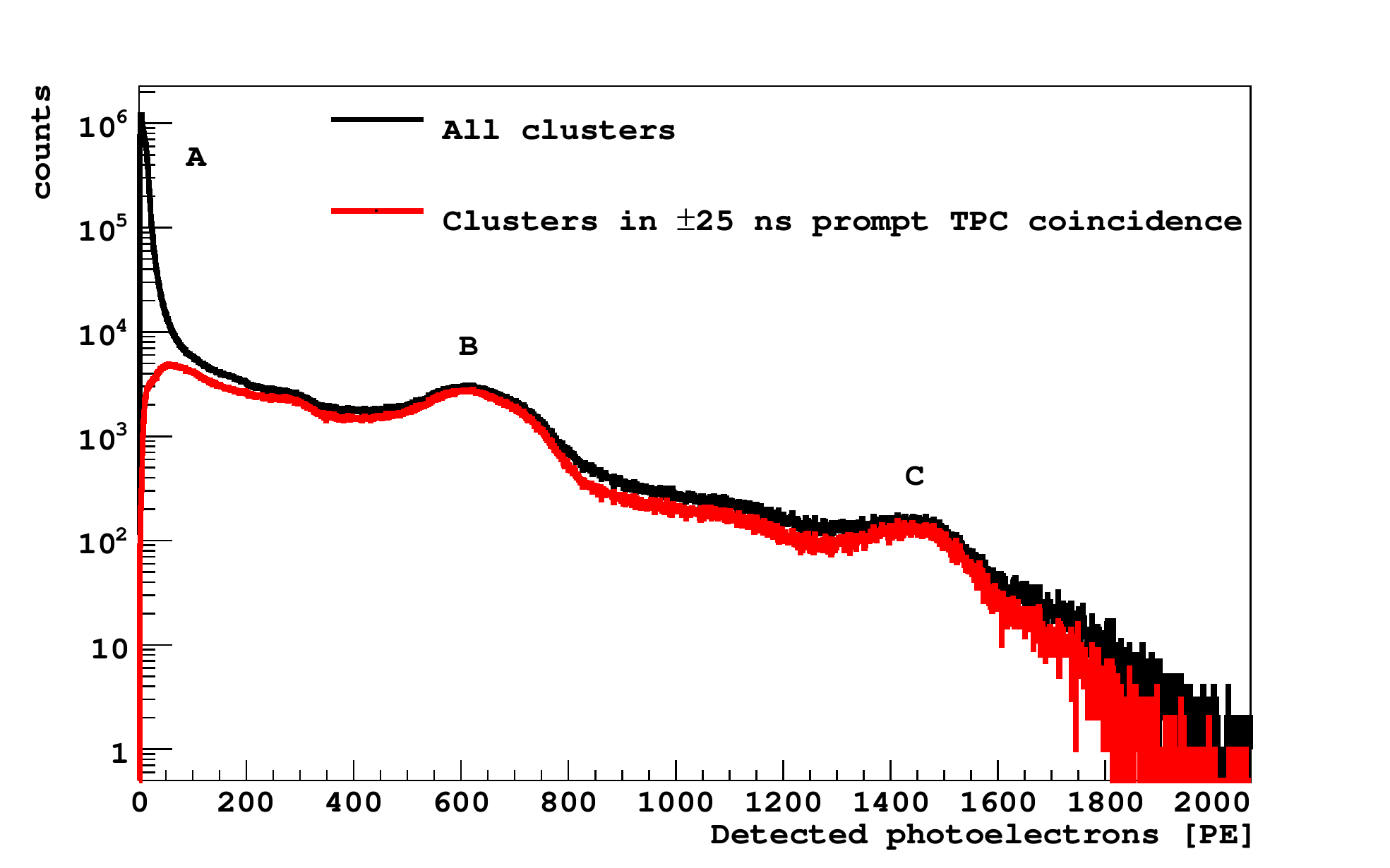}
 \caption{Photoelectron spectrum of the \lsv\ clusters during \phasetwo: (black) for all clusters and (red) for clusters starting within a $\pm$25 ns prompt coincidence with the \lar\ \tpc. Region A is the low energy signal that consists largely of \aftp s, Region B is the \coba\ peak, and Region C is the \thal\ peak.}
 \label{fig:fullespec}
\end{figure}

Crucial aspects of the \lsv\ performances include the rates of various backgrounds, and the light yield (the number of PE collected by all of the \pmts\ per keV deposited in the scintillator).
Measurements of the background rates and the light yield of the \lsv\ are made by building a photoelectron spectrum from the clusters and fitting it with the simulated spectra from known radioactive contaminants in the materials of the detector.
The shape of the radioactive background includes the detector response function, in which the light yield is a parameter. 
This method is complementary to efforts to calibrate the \lsv\ light yield through the deployment of sealed \gr\ sources within the detector.

The most prominent radioactive contaminants visible in the \lsv\ are \cfor, \coba\ and \thal.
The full photoelectron spectrum of the \lsv\ is shown in figure~\ref{fig:fullespec}.
The light yields derived from fits to the \cfor\ $\beta^-$ spectrum, the two \coba\ \gr s, and the \thal\ \gr\ in \phasetwo\ are summarized in table~\ref{table:ly}. The calculation of the light yield from the \cfor\ $\beta^-$ spectrum and the \coba\ \grs\ assume a quenching model described by a Birks' constant $kB\,=\,0.012$\,cm/MeV~\cite{Bellini:2014ke}. 

\begin{table}[tb]
 \caption{Light yield measurements during \phasetwo\ from fits to four different background sources in the \lsv\ photoelectron spectrum. Errors given here are statistical from the respective fits.}
 \centering
 \begin{tabular}{l|ccc}
 \hline
  Isotope & Decay mode & Energy (keV)   & Light Yield (PE/keV)\\
  \hline \hline
  \cfor    & $\beta^-$           & 156 (endpoint)& 0.561$\pm$0.013\\
  \coba    & $2\gamma$           & 1173, 1332     & 0.592$\pm$0.011\\         
  \thal    & $\gamma$            & 2614           & 0.551$\pm$0.002\\
  \hline
 \end{tabular}
 \label{table:ly}
\end{table}

\subsubsection{Low photoelectrons background}
\label{sec:low-energy}
In the region labeled A in figure~\ref{fig:fullespec}, we observe a class of events with very low photoelectron yield and a high rate. 
Most of these events appear to be very highly concentrated in a small number of \pmts, though the \pmts\ contributing to each pulse vary.
Investigation of these events revealed that a large fraction of them are due to \aftp s. This is supported by the fact that the rate of these events is significantly higher in the 10-20\,$\mu$s following a scintillation event, while it is significantly diminished in the time window before the scintillation.
However, \aftp s do not appear to account for the entire structure.
It was also found that the rate of the low photoelectrons events spiked whenever fluid operations were performed, but that the spike would diminish down to the pre-operations rate on a timescale of a few days.
One possible explanation is that the charge may build up in the electrically-insulating \pc\ when it moves through the metal pipes and then discharges on individual \pmts.

\subsubsection{\cfor\ background}
\label{sec:c14}

\begin{figure}[tpb]
\centering
\includegraphics[width=0.8\linewidth]{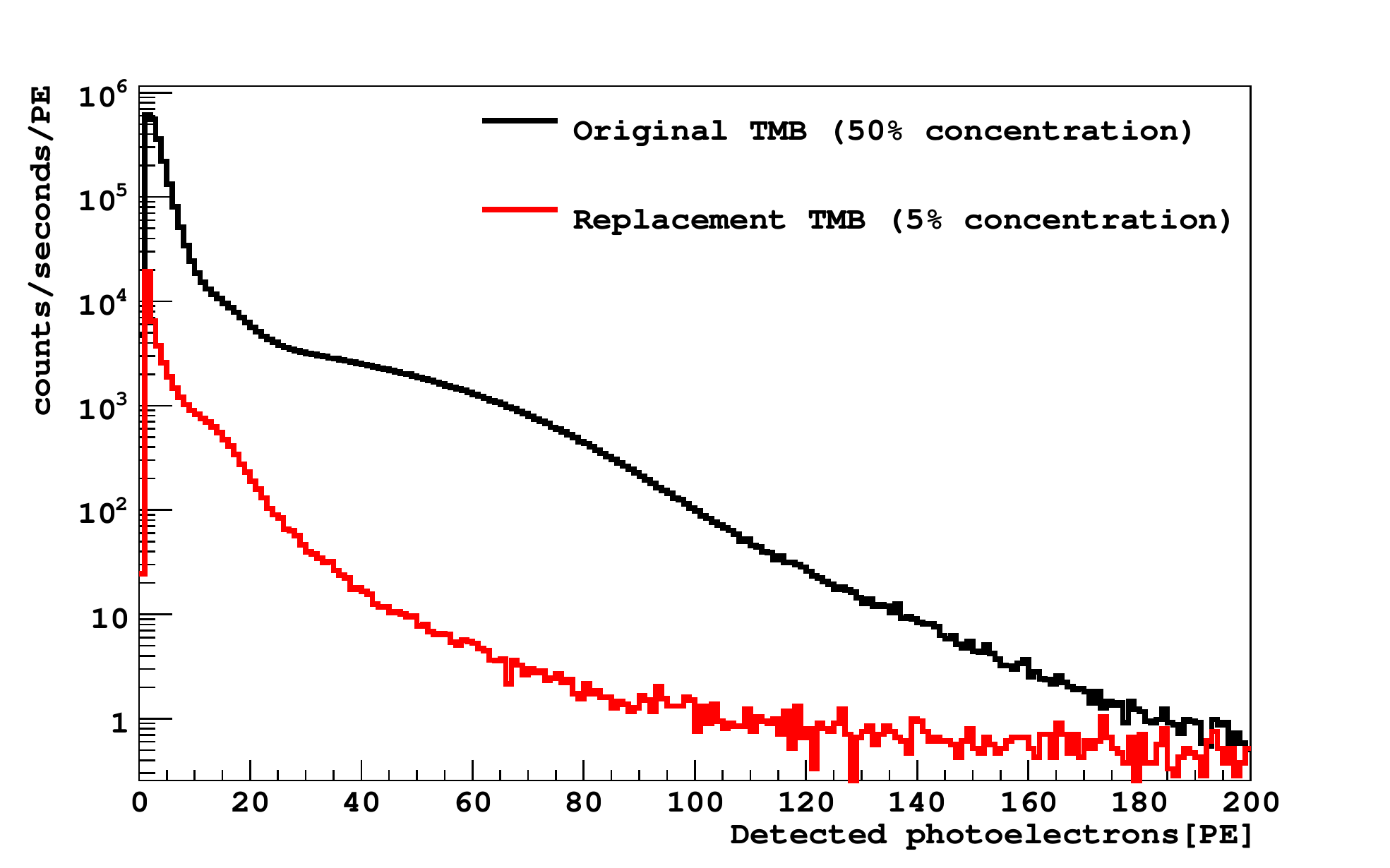}
\caption{ 
The low end of the photoelectron spectrum seen in the \lsv\ in \phaseone, with \cfor-rich \tmb, at 50\% concentration, compared to the \phasetwo\ low-\cfor\ \tmb\ at $\sim5\%$ concentration. The dramatic decrease in \cfor\ background is evident.}
 \label{fig:veto14c}
\end{figure}

\begin{figure}[tb]
\centering
 \includegraphics[width=0.8\linewidth]{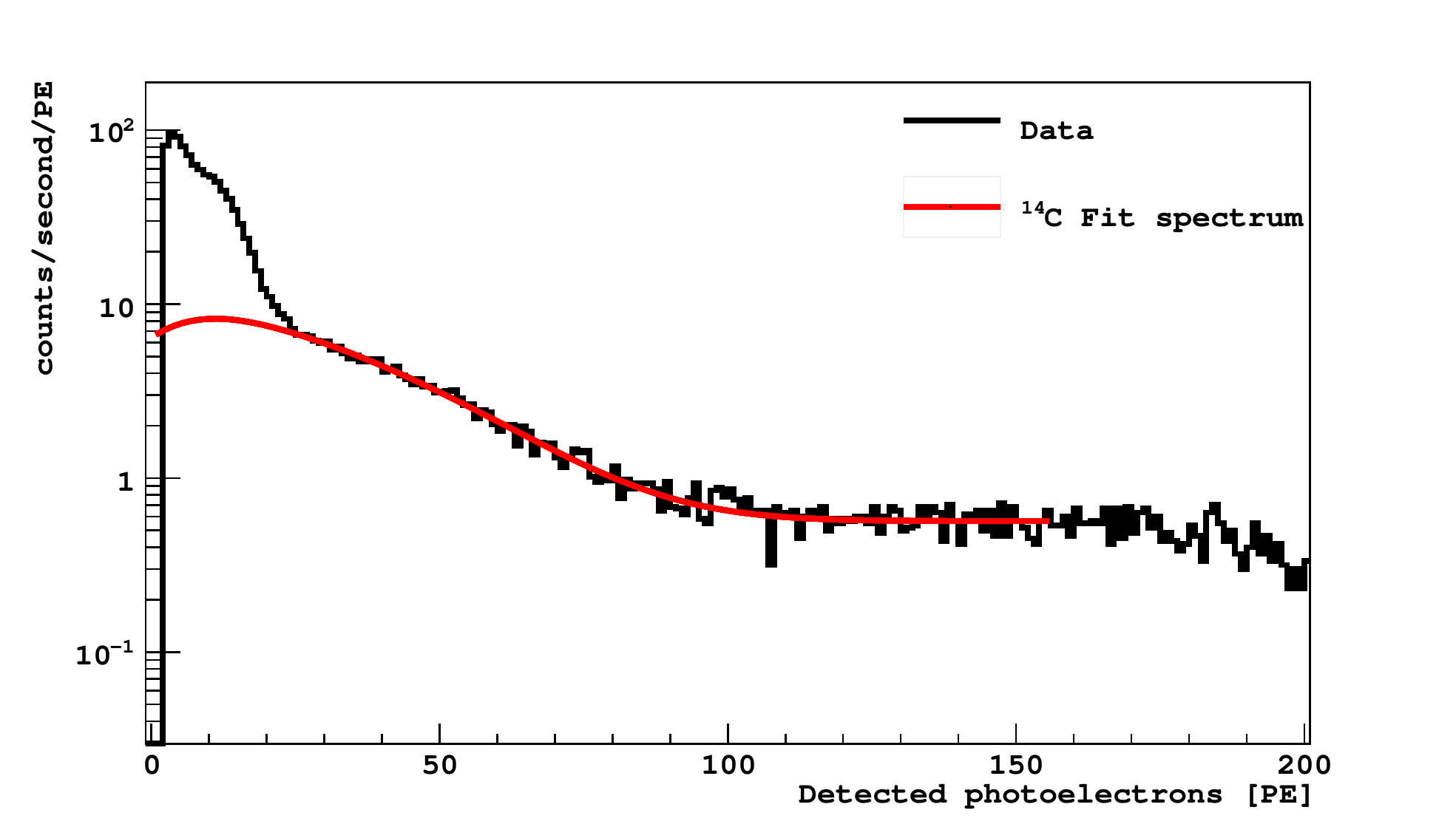}
\caption{Photoelectron spectrum of the \cfor\ in the \lsv\ in \phasetwo, evaluated with the clustering method. The black curve is the measured low energy spectrum after applying cuts to remove \aftp s, and the red curve is the theoretical \cfor\ spectrum, assuming a constant shape function, convolved with a Gaussian response function with light yield, resolution, background (assuming a constant background in this energy region), and \cfor\ rate as fit parameters. Fitting this curve from 25--120\,PE gives a light yield of 0.561\,$\pm$\,0.013\,PE/keV, and a \cfor\ rate of \lsvcforratelow.}
\label{fig:c14}
\end{figure}

The \cfor\ isotope is present in any organic liquid like \pc\ and \tmb, and it 	is therefore an unavoidable radioactive background. 
\cfor\ decays by $\beta^-$ emission with a Q-value of 156\,keV, giving a background concentrated at the low energy part of the \lsv\ spectrum.

The original batch of \tmb\ used in the \lsv\ during \phaseone\ (discussed in section~\ref{sec:scint}), used \tmb\ derived primarily from modern biogenic methanol, and  had a \cfor\ contamination of \lsvcforratehigh.
This activity is only a factor ten less than what would be expected from a sample of modern atmospheric carbon of equal mass ($\sim$15\,tonnes).
In order to increase the sensitivity of the \lsv, the \tmb\ was replaced with a new batch, produced by a petroleum-based process, with substantially lower \cfor\ contamination for \phasetwo.
The process of removing the old \tmb\ and replacing it with the new batch is described in section~\ref{sec:fluidhandling}.
Figure~\ref{fig:veto14c} compares the \cfor\ spectra with the original \cfor-rich \tmb, at 50\% concentration and the current low-\cfor\ \tmb\ at $\sim5\%$ concentration. 

Figure~\ref{fig:c14} shows the $\beta^-$ spectrum of \cfor\ convolved with the fitted \lsv\ response,
compared to the \phasetwo\ data. The fitting function includes the energy resolution and ionization quenching (assuming a Birks' constant of 0.012\,cm/MeV~\cite{Bellini:2014ke} and the model described in~\cite{GrauMalonda:1999ex}), and an additional constant background term. 
The fit is performed over the photoelectron distribution of clusters in random coincidence with the \lar\ \tpc\ between 25 and 120\,PE. Below 25\,PE the low photoelectron background is dominant (see section~\ref{sec:low-energy}) and thus we exclude this region.
The region above 100\,PE is dominated by the constant term.
To reduce events due to of \aftp s, only clusters that followed a window of at least 30\,$\mu$s with no clusters were considered.
The \cfor\ decay rate measured during \phasetwo\ is \lsvcforratelow. 

Borexino, which uses \pc\ from the same source as \dsf, measured their petroleum-derived \pc\ to have a \cfor\ contamination of 40\,Bq per 100\,tonnes~\cite{Bellini:2014dl}.
Normalizing for mass, the Borexino liquid scintillator has a \cfor\ specific activity of $\sim$0.4\,mBq/kg, while the \phaseone\ \dsf\ liquid scintillator had $\sim$5000\,mBq/kg, and the \phasetwo\ \dsf\ liquid scintillator has $\sim$8.3\,mBq/kg.
The latter specific activity is consistent with a residual contamination by $\sim15\,$kg of the biogenic \tmb,
assuming the petroleum-derived \tmb\ has a comparable \cfor\ contamination to the petroleum-derived \pc.

\subsubsection{\coba\ background}
\begin{figure}[tb]
\centering
 \includegraphics[width=0.8\linewidth]{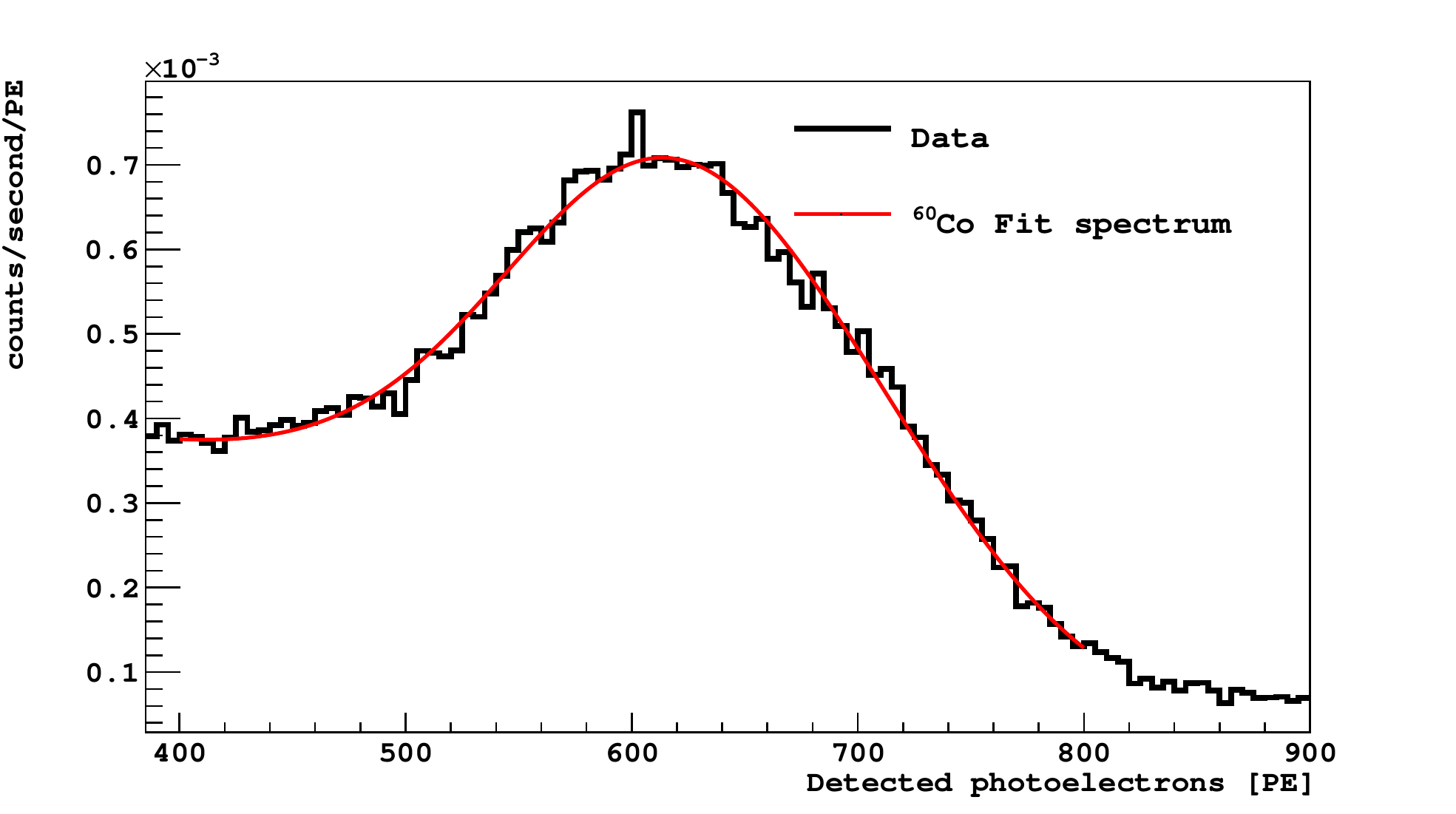}
\caption{Photoelectron spectrum of the \coba\ in the \lsv\ in prompt coincidence with a signal in the \lar\ \tpc\ (black), compared to a fit of a simulated energy distribution convolved with a Gaussian response function (red). The fit, performed around the two peaks from the 1.17 and 1.33\,MeV \grs, gives a light yield of 0.592$\pm$0.011\,PE/keV.}
\label{fig:co60}
\end{figure}

The \coba\ isotope is an anthropogenic isotope found in most samples of stainless steel, primarily as a result of uncontrolled \coba\ disposal, leading to the isotope getting mixed in with recycled stainless steel, and from \coba\ added to the steel during the production process~\cite{maneschg_measurements_2008}.
This isotope is therefore present in the stainless steel of both the sphere of the \lsv\ and the \lar\ \tpc\ cryostat. Measurements using germanium counting techniques indicate that the cryostat stainless steel has a specific activity of 13.1$\pm$1\,mBq/kg of \coba, while the \tpc\ \pmts\ have an activity of 8.8$\pm$0.8\,mBq/PMT, and the \lsv\ stainless steel has 2.8$\pm$0.3\,mBq/kg. 

\coba\ $\beta^-$ decays with a half-life of $5.3$~years to {\mbox{$^{60}$Ni}}.
The excited nickel nucleus promptly emits two \grs\ with energies of $1.17$ and $1.33$\,MeV. 
Since one of the \grs\ may interact in the \lar\ \tpc\ while the other interacts in the \lsv, the probability of seeing prompt coincidences with peaks at both \gr\ energies is greatly enhanced.
Prompt coincidences between the \lar\ \tpc\ and the \lsv\ are, in fact, dominated by 
 \coba\ in the \lsv\  above 1\,MeV.
A  sample of \gr s from \coba\ events can therefore be obtained in the data by selecting scintillation events in the \lsv\ in a narrow time coincidence with a scintillation event in the \lar\ \tpc.

The \coba\ peak is shown in the region labeled B in figure~\ref{fig:fullespec}. 
Figure~\ref{fig:co60} shows the \gr\ peak of \coba\ decay in the \lsv\ spectrum in coincidence with the \lar\ \tpc. A fit was performed by simulating \coba\ decays in the cryostat stainless steel and \pmts\ using \geant, and then convolving the simulated distribution of energy deposited in the \lsv\ with a Gaussian response function, similar to the one used for \cfor. The results from this fit are also shown in figure~\ref{fig:co60}. This fit gives a light yield of 0.592$\pm$0.011\,PE/keV. Due to the finite resolution of the \lsv, the peaks from the two \grs\ are not clearly distinguishable.

\subsubsection{\thal\ background}
\begin{figure}[tb]
\centering
 \includegraphics[width=0.8\linewidth]{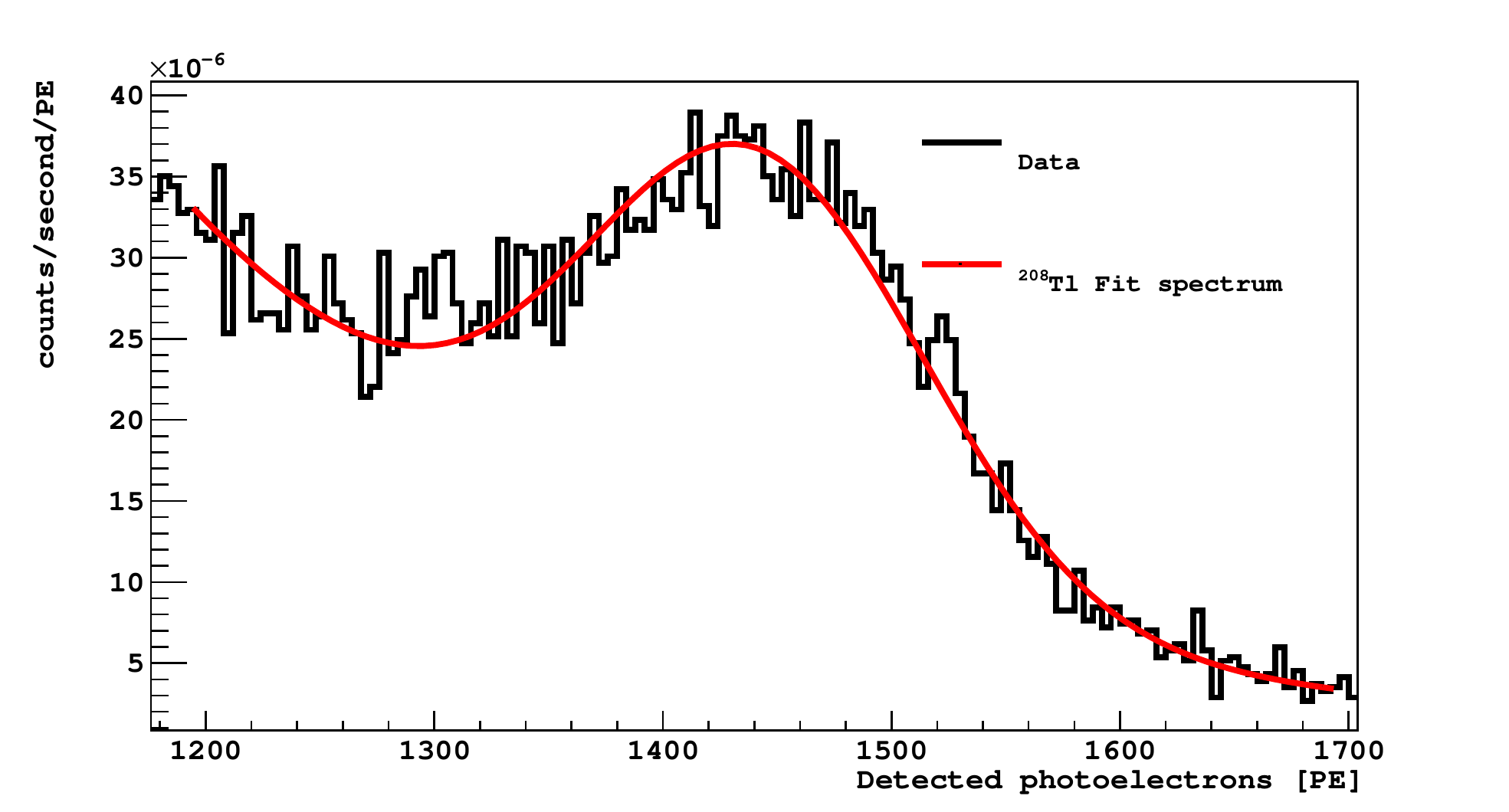}
\caption{Photoelectron spectrum of the \thal\ in the \lsv\ in prompt coincidence with a signal in the \lar\ \tpc\ (black), compared with a fit to a Gaussian plus a falling exponential fit to it (red). The 2.6\,MeV \gr\ peak is seen at 1440\,$\pm$\,5\,PE.}
\label{fig:tl208}
\end{figure}
\thal\ has a half-life of 3.05 minutes and is generated in the naturally-occurring \tho\ decay chain, which is naturally present in most samples of iron due to its abundance in the earth's crust and its half-life of $\sim$14 billion years. 

\thal\ $\beta^-$ decays with a Q-value of 5\,MeV. Around 99.8\% of these decays are also accompanied by a 2.6\,MeV \gr. Additionally, 85\% are accompanied by a 583\,keV \gr, as well as a large number of other potentially coincident \grs. The 2.6\,MeV \gr\ is the highest energy \gr\ expected from the uranium and thorium decay chains. This high energy, in addition to its coincident \gr s that may trigger the \lar\ \tpc\ while it goes into the \lsv, makes the full energy peak easily identifiable, as can be seen in the region labeled C in figure~\ref{fig:fullespec}.

Figure~\ref{fig:tl208} shows the full energy peak of \thal\ with a Gaussian plus an exponential background fit to it. The mean of the Gaussian, corresponding to the 2614\,keV full energy peak, is at 1440$\pm$5\,PE, corresponding to a light yield of 0.551$\pm$0.002\,PE/keV.

\subsection{\bifo-\pofo\ and \radon}
\label{sec:bipo}
\bifo\ and \pofo\ are two isotopes that decay in sequence in the \radon\ chain.
\bifo\ decays by $\beta$ or $\beta + \gamma$ emission with a Q-value of $3.23$\,MeV.
\pofo\ decays via $\alpha$ emission with a very short lifetime ($\tau \, = \, 236.6\, \mu$s); the energy of the $\alpha$ is $7.7$\,MeV~\cite{Bellini:2013dp}.

The short lifetime of \pofo\ makes it possible to estimate the activity of \bifo, \pofo, and therefore of \radon\ in the \lsv, by studying the time distribution of consecutive scintillation events.
The data used to evaluate the \bifo-\pofo\ contamination were taken after the \tmb\ removal (see section~\ref{sec:lsv_reconstitution}), in \lsv\ self-trigger configuration (see section~\ref{sec:reconstruction}).
The majority trigger was generated if at least 20 \pmts\ of the \lsv\ fired in a window of 60\,ns.
The acquisition window was set to $6.5$\,$\mu$s to acquire the scintillation signal and a sample of the \pmt\ after-pulsing following the scintillation.
The time delay between consecutive scintillation events in the \lsv\ shows a correlated time distribution with a lifetime in good agreement with the one expected from the \pofo\ decay.

The photoelectron spectra of the candidate \bifo\ and \pofo\ events, after subtraction of random coincidences, are compatible with the expected decay spectra.
By integrating the number of consecutive events in the \bifo-\pofo\ energy window, a \radon\ activity of $\sim1.5$\,mBq or 44\,counts-per-day$/$tonne is estimated in the \lsv. The background due to \radon\ is therefore several orders of magnitude less than the backgrounds due to \cfor\ or \coba.

The measurement of the scintillation light emitted in the \pofo\ decay allows an evaluation of the quenching factor of $\alpha$ particles at $7.7$\,MeV, at different \ppo\ concentration in the \lsv. 
The \pofo\ $\alpha$ peaks, with no \tmb\ and at \ppo\ concentrations of $\leq 0.1$\,g$/$L and \lsvppoconcorig, respectively, are quenched by a factor of $\sim12$ and $\sim8$.
We recall that the scintillation yield from low-energy $\alpha$ particles is quenched more heavily than is the yield from higher energy ones. This means that the low-energy $\alpha$ particles produced by neutrons capturing on \borten\ are quenched more heavily than those produced by the $\alpha$ decay of \pofo.

\subsection{Muons in the veto}
\label{sec:reco-muons}

\begin{figure}[tb]
 \centering
 \includegraphics[width=0.8\linewidth]{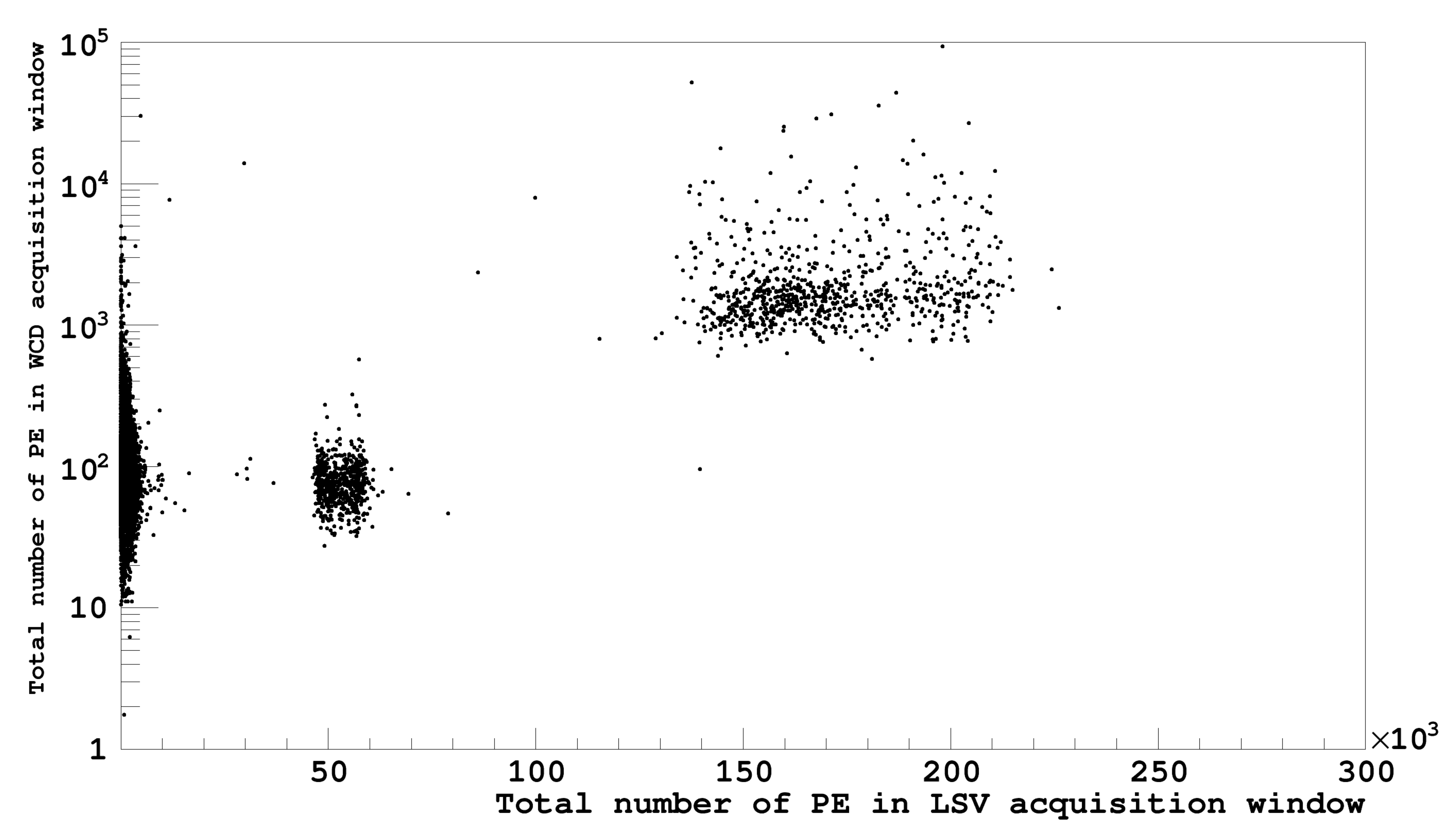}
 \caption{The total number of photoelectrons detected in the \wcd\ and \lsv\ during \phasetwo, with the \lsv\ and \wcd\ triggering on the \lar\ \tpc. Below 10$\times10^3$~PE in the \lsv\ are normal scintillation events in the \lsv\, mostly from \grs, and \pmt\ noise in the \wcd. Between $40\times10^3$ and $70\times10^3$~PE in the \lsv\ are events with large electronics noise in the \lsv. At higher charges are events in which muons passed through the \wcd, the \lsv, and the \lar\ \tpc. }
 \label{fig:muon_id}
\end{figure}

Both the \lsv\ and the \wcd\ can identify muons.
Due to the high energy deposited (about 2\,MeV$/$cm), muons can clearly be identified in both the \lsv\ and the \wcd\ by looking at the total amount of energy deposited in either detector for the entire acquisition window, as shown in figure~\ref{fig:muon_id}.
A study of muons in both detectors shows that muons can be very efficiently identified in the \lsv\ by looking for a total charge above 2000\,PE, and in the \wcd\ by looking for a total charge above 400\,PE.
In both cases, the background contributions due to noise or event pile-up are $<0.1\%$.

The rate of muons crossing both the \lsv\ and the \wcd\ was found to be $380\, \pm \,5$\,day$^{-1}$, in good agreement with the expected rate of 370\,muon$/$day extrapolated from the measurement of the muon rate in the Hall C of LNGS performed in Borexino~\cite{bellini_cosmic-muon_2012}.

In the dark matter analysis, the muon rejection is done using both \lsv\ and \wcd\ capabilities.
The detailed study of the overall muon tagging efficiency and cosmogenic neutron rejection power in \dsf\ will be reported in a future paper.

\subsection{\lsv\ calibration with \AmBe\ source}
\label{sec:reco-ambe}

\begin{figure}[tb]
\centering
 \includegraphics[width=0.8\linewidth]{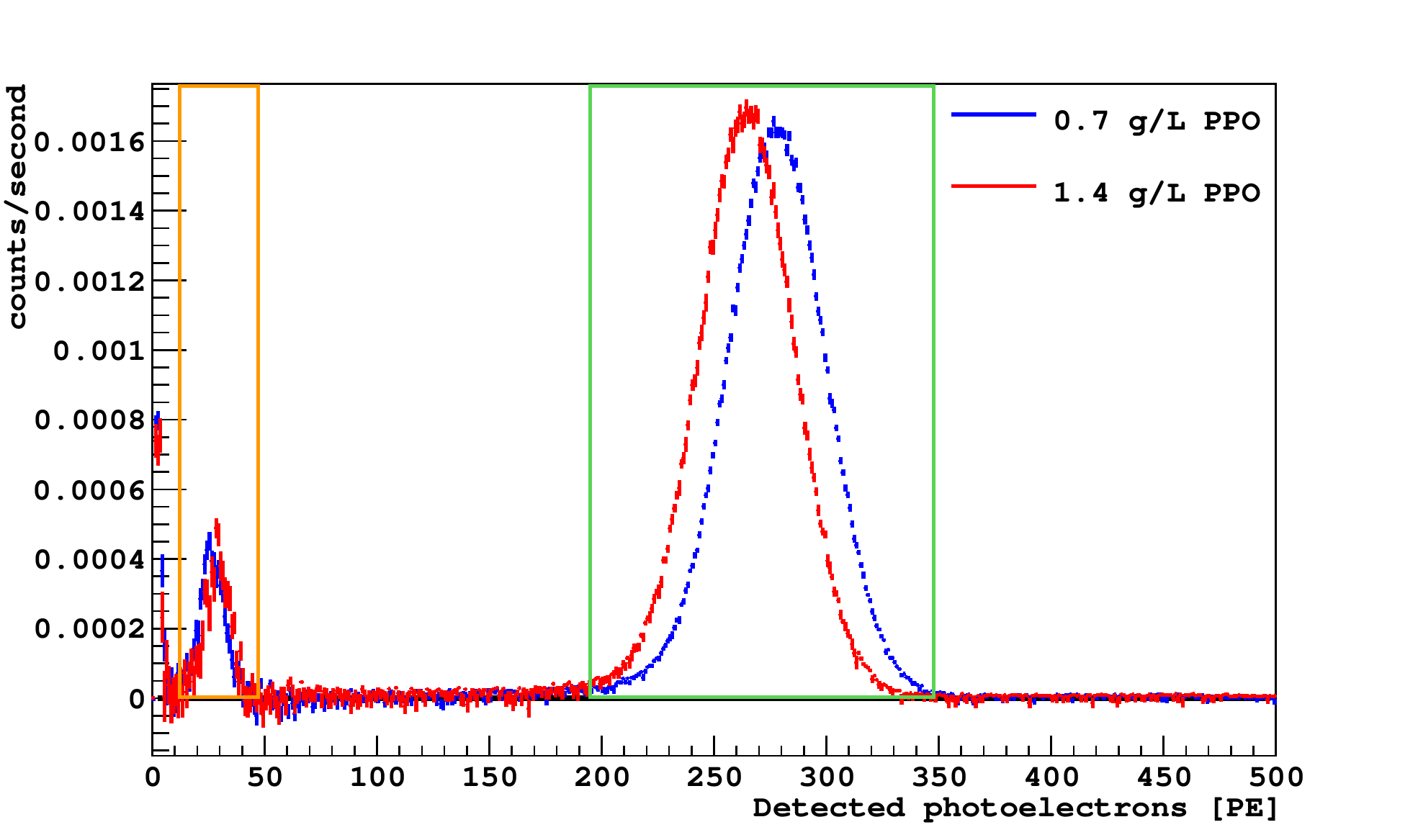}
\caption{
The photoelectron spectrum of neutron captures in the \lsv\ at 5\% \tmb\ and \lsvppoconcint\ \ppo\ (blue) and at  \lsvppoconcnow\ \ppo\ (red), for signals in delayed coincidence with a large \lsv\ signal from the $4.4$\,MeV \gr\ accompanying many \AmBe\ neutrons, after background subtraction. The peak on the left at $\sim30$\,PE (orange box) is from the
\brbortenground\ of neutron captures on \borten\ which lead to an $\alpha$ + \lith\ ground state,
while the peak on the right at $\sim270$\,PE (green box) is from the \brbortenexcited\ of captures that lead to the \lith\ excited state reaction,
with the accompanying \enbortenexcitedgamma\ \gr.
The features below 10\,PE are due to \pmt\ \aftp s. These measurements were taken with the \AmBe\ source $\sim$70\,cm away from the cryostat.
}
\label{fig:ambe_spectrum}
\end{figure}

\begin{figure}[tb]
\centering
 \includegraphics[width=0.8\linewidth]{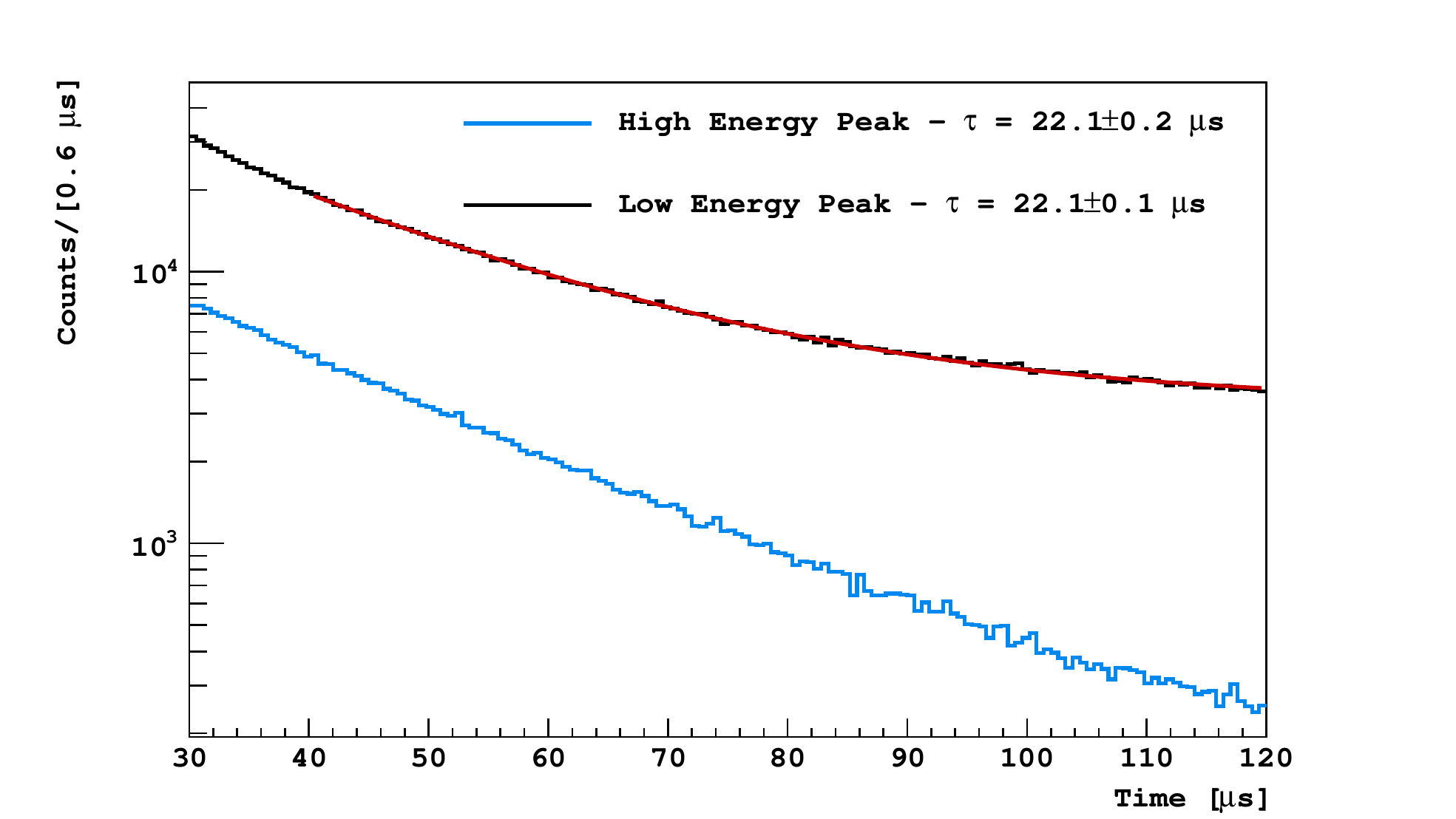}
 \caption{The time distribution for neutron captures measured during an \AmBe\ calibration campaign. The two curves correspond to the low energy (black) and high energy (blue) peaks, corresponding to the neutron captures on \borten\ leading to the \lith\ ground state and excited state, respectively. An exponential plus a constant were fit to both of these curves (shown in red in the low energy peak curve), giving a decay time of $22.1$\,$\mu$s, consistent with the expected capture time for thermal neutrons during \phasetwo. The low energy peak appears to have a higher rate than the high energy peak in this plot because it is offset by a higher random background rate, which is accounted for by the constant term in the fit.}
 \label{fig:ambe_capture_time}
\end{figure}
The \AmBe\ neutron source deployment with CALIS (see section~\ref{sec:CALIS}) allowed an \emph{in-situ} calibration of the response of the \lsv\ to neutron captures.
During the \AmBe\ calibration the \tmb\ concentration in the \lsv\ was  5\%.
The \AmBe\ source had a neutron activity of 10 neutrons per second.
In order to study the position dependence of the light output, we deployed the \AmBe\ source in two different positions, one in physical contact with the \lar\ \tpc\ cryostat, and one $\sim72$\,cm away from the cryostat.
Data was taken both with the \lsv\ trigger driven by the \lar\ \tpc\ trigger, and self-triggered on the signal from the $4.4$\,MeV \gr\  that accompanies a large fraction of \AmBe\ neutrons~\cite{kamboj_precise_1986}.
We performed \AmBe\ calibrations at two different \ppo\ concentrations: \lsvppoconcint\ and \lsvppoconcnow,
allowing a characterisation of the response of the \lsv\ to neutrons and a direct measurement of the impact of the \ppo\ concentration on the $\alpha$ quenching. 

In order to measure the energy spectrum of the neutron capture products without being biased by \aftp s from the prompt signal (due to the neutron thermalizing and to coincident \grs\ from the source), the first $20\,\mu$s after the prompt signal is ignored. 
The photoelectron spectrum of the first cluster identified after this period for each event is shown in figure~\ref{fig:ambe_spectrum}, with random backgrounds measured with a background run  subtracted. The lowest energy feature in figure~\ref{fig:ambe_spectrum} consists of \aftp s from the prompt signal, 
i.e. a neutron captures in the first $20\,\mu$s so we miss the signal, but we detect the \aftp\ and misidentify it as a neutron capture signal.
The low energy peak, which we associate with the \lith\ ground state channel, lies between 20 and 30\,PE, well above our detection and analysis threshold in \phasetwo\ (see section~\ref{sec:slider}).
We found that increasing the \ppo\ concentration increased the light yield for the $\alpha\,+^7$Li channel by 3 PE ($\sim$11\% increase) and reduced the $\alpha\,+^7\text{Li}\,+\,\gamma$ peak light yield by 13 PE ($\sim$5\% decrease). The decrease in the peak position of the latter channel is likely either due to increased self-absorption at the higher \ppo\ concentration or to contaminants inadvertently added to the scintillator along with the \ppo. The increase in the peak position of the former channel, despite the decrease in the light yield of the latter, is likely due to the higher \ppo\ concentration decreasing the effects of ionization quenching.

To verify that the two peaks in figure~\ref{fig:ambe_spectrum} are from neutron captures on \borten, we studied the time distributions of the clusters relative to the trigger
(either the $4.4$\,MeV \gr\ in the \lsv\ or a neutron-like event in the \lar\ \tpc).
The clusters of both peaks follow an exponential distribution with a time constant of 22\,$\mu$s, as shown in figure~\ref{fig:ambe_capture_time},
consistent with the capture time expected from \geant\ simulations.
The spectrum and time distribution of the ground state channel shows a non-negligible constant background term,
mainly due to \pmt\ \aftp s following the $4.4$\,MeV \grs.

The limited reflectivity of the polished stainless steel \tpc\ cryostat leads to the loss of some of the scintillation light which strikes it.
The reconstructed energy of the neutron capture signals therefore depends on the position of the source relative to the \lar\ \tpc.
This effect is visible in the the neutron capture peaks, 
when the \AmBe\ source is moved from a position $\sim$72\,cm away from the \lar\ \tpc\ cryostat to a position in physical contact with it.
The light yields of the $\alpha\,+^7$Li channel and the $\alpha\,+^7\mbox{Li}\,+\gamma$ channel, respectively, decrease by $\sim26\%$ and $\sim10\%$. 
This is in agreement with \geant\ simulations of the \lsv, which predict that the light yield of a scintillation event very close to the \tpc\ cryostat should decrease by $\sim30\%$ with respect to an event in the bulk of the \lsv.	

The 478\,keV \gr\ of the $\alpha\,+^7\mbox{Li}\,+\gamma$ channel of \borten\ capture can escape the \lsv\ by going back into the cryostat, with a probability of $\sim8\%$, as predicted by \geant\ simulations. If this happens, the scintillation light will be produced only by the 1471\,keV $\alpha$ and the 839\,keV $^7\mbox{Li}$. The energy deposited in the \lsv\ in the former case is $\sim82\%$ of the energy deposited in the ground state reaction, where a 1775\,keV $\alpha$ and a 1015\,keV $^7\mbox{Li}$ are produced. The energy resolution at 20-30\,PE does not allow discrimination between the two cases.
The neutron capture on \borten\ producing an excited state \lith, when the 478\,keV \gr\ escapes the \lsv, is therefore reconstructed as a capture producing a ground state \lith.
 
The predicted 8\% probability of losing the 478\,keV \gr\ was confirmed in the \AmBe\ data by looking at data collected with the source near and far from the cryostat. With the source far from the cryostat, we found that $\sim6\%$ of neutrons went to the $\alpha$+\lith\ channel, consistent with the expected branching ratio of $6.4\%$. However, when the source was rotated near the cryostat, we found that $\sim14\%$ of the neutrons resulted in a capture signal in the $\alpha$+\lith\ peak. Since we cannot resolve the energy differences between the $\alpha$ and \lith\ produced by both capture channels, this is consistent with 8\% of the 478\,keV \grs\ being lost, making the captures that go to \lith$^*$ look the same as those that go straight to the ground state.

\geant\ predicts that $\sim8\%$ of neutrons capture on \hydrogen, producing a $2.2$\,MeV \gr. Additionally, \geant\ predicts that approximately $8\%$ of these \grs\ do not leave a visible signal in the \lsv, leaving the neutron capture undetected.
We used \geant\ to predict the ratios of the \emph{reconstructed} \hydrogen\ captures and \borten\ producing excited state \lith\ captures to the \borten\ producing ground state \lith\ captures, considering the effect of escaping \grs, in the \AmBe\ calibration.
The ratios are compatible with the ones measured from \AmBe\ data, by counting the number of events under the capture peaks in the  background subtracted spectrum.

In summary, there are five main channels through which we may lose the detection of a neutron capture.
\begin{enumerate}
\item Neutron captures on \hydrogen, and the 2.2\,MeV \gr\ escapes the \lsv\ without leaving a visible signal\label{enum:hlosschan}.
\item Neutron captures on \borten, which goes to the ground state and produces a signal below the detection threshold\label{enum:borgroundlosschan}.
\item Neutron captures on \borten, which goes to the excited state, but the 478\,keV \gr\ escapes and the resulting signal from the $\alpha$ and \lith\ is below the detection threshold\label{enum:borexcitedlosschan}.
\item Neutron captures before it reaches the \lsv\ and the capture does not result in a \gr\ detected by the \lsv.\label{enum:nolsvlosschan}
\item Neutron captures after the acquisition \lsv\ window has ended.\label{enum:latelosschan}
\end{enumerate}

Given the low threshold we place on the delayed coincidence cut, we expect the loss in efficiency due to channels \ref{enum:borgroundlosschan} and \ref{enum:borexcitedlosschan} to be $<\,0.1\%$. \geant\ predicts that  $\sim0.64$\% of neutron captures are undetectable due to channel~\ref{enum:hlosschan}. Additionally, as mentioned in section~\ref{sec:neutron_detection}, $\sim0.05$\% of neutrons are expected to be lost through channel~\ref{enum:nolsvlosschan}. Lastly, the thermal neutron capture time constant compared to the length of the acquisition window during \phasetwo\ predicts that $\sim0.23\%$ of neutron captures will be lost through channe~\ref{enum:latelosschan}.
Consequently the efficiency of the \lsv\ for detecting radiogenic neutron captures is at least 99.1\%. The neutron detection efficiency could increase by detecting the scattering of neutrons on protons and carbon nuclei during thermalization.

\subsection{Detector stability}
\label{sec:stability}


The long-term stability of the detectors has been studied using three metrics:
\begin{enumerate}
 \item \emph{Light yield stability}: In order to run for a long time, it is necessary that the light yield (which depends on the reflector, \pmt, and scintillator stability) remain high throughout the duration of the dark matter search campaign. The light yield stability is also important for placing cuts that will maintain their effectiveness over time. This was monitored by studying the movement of the easily identifiable \coba\ peak over time. During \phaseone, which went from Nov. 6, 2013 to Jun. 3, 2014, the light yield increased with an average slope of 
0.3\,$\pm$\,0.1\,PE/week. During \phasetwo, which went from Apr. 8, 2015 to Jul. 31, 2015, the light yield increased with an average slope of 
0.2\,$\pm$\,0.1\,PE/week. The causes are still under investigation, nevertheless, over the course of the runs, this change in light yield was found to be small enough so that it can be neglected.

  \item \emph{\lar\ \tpc\ coincidence timing stability}: In order to ensure that prompt cuts consistently identify coincidences between the veto and the \lar\ \tpc, it is important to monitor the stability of the  timing with respect to the primary scintillation signal in the \lar\ \tpc, which may vary with any sort of changes made to the electronics and DAQ. Possible timing offsets are monitored by looking at the distribution of the start times of clusters.
 True \lsv-\tpc\ coincidences typically give a sharp timing peak, and so cluster start times can be used to track the relative timing of with an accuracy $<$\,10\,ns. 
 The \lar\ \tpc\ coincidence timing remained stable throughout the duration of the dark matter search campaign.
 
 \item \emph{Event rate stability}: To monitor for changes in the signal that the \lsv\ detects, it is important to track  the event rate of the detector. With the exception of a minor change in the data acquisition system made about a fifth of the way through \phasetwo, which caused an increase in the event rate by $\sim$1.2\%, the event rate was stable to within 0.03\% per week over the course of the 70 day campaign. It should be noted that the change in the data acquisition system corresponds to a fifth of the runs in the campaign, but about half of the livetime used for the dark matter search in \phasetwo~\cite{Agnes:2015_uar}.
\end{enumerate}

Spikes at low energy (below $\sim$20\,PE, see section~\ref{sec:low-energy}) were observed following operations in the fluid, such as those involved with the reconstitution of the scintillator. However, the spikes diminished to a baseline level within a few hours of the operations ending. These low energy and high rate events were seen to be highly concentrated in a small number of phototubes, with an unusually fast time signature. One possible explanation is that the motion of the scintillator during the operations caused charge to build up, which then discharged on the $\mu$-metal grids. A slower (on the time scale of $\sim$150 days) decrease in this event rate was also observed, which has eventually stabilized at a constant background rate of $\sim$1\,kBq after making time cuts to remove \aftp s from prompt scintillation events.

It is worth noting that the optical properties and stability of the \lsv\ were not affected by the two calibration campaigns and no additional radioactivity was introduced by the CALIS system.

\subsection{Summary of estimated neutron rejection power}

\geant\ simulations suggest that 99.95\% of neutrons that undergo a single nuclear recoil in the WIMP search region in the \lar\ \tpc\ of \dsf\ deposit some energy in the \lsv.
The neutron rejection power is thus based upon the ability to detect the signals produced by the neutron interactions in the \lsv,
namely from the scattering on protons and \ctwe\ during thermalization and from the neutron capture.
At the reconstruction and analysis levels, the information about these two signals is contained respectively in the prompt region of interest (section~\ref{sec:roi}) and the sliding window (section~\ref{sec:slider}) variables.
In the end, the neutron rejection power depends on the thresholds and  time windows used for these variables.
A gain in the neutron capture rejection power is possible at the cost of increasing the probability of vetoing on random \lsv\ backgrounds, thus decreasing the acceptance of the cuts in the WIMP search.

In \dsf\ \phaseone~\cite{Agnes:2015gu}, the neutron rejection power was limited due to the \cfor\ background in the \lsv.
However, the small neutron capture time ($\tau\, \sim 2.2\,\mu$s) due to the high concentration of \tmb,
allowed us to set a neutron detection efficiency of $\sim98\%$ with an acceptance loss of 11\%, including an approximation of the thermalization signal rejection power based on~\cite{hong_scintillation_2002}.

In \dsf\ \phasetwo~\cite{Agnes:2015_uar}, the low background level in the \lsv\ allows for a much higher neutron detection efficiency, even at a low concentration of \tmb\ (5\%), by combining  the detection of the neutron capture and the thermalization signals.
The neutron rejection power for the neutron capture signal alone is $\geq\,99.1$\%.
The main limit on the neutron capture efficiency is due to residual neutron captures on \hydrogen, happening in $\sim8$\% of cases.
Simulation and \AmBe\ calibration data indicate that $\sim8$\% of the $2.2$\,MeV \grs\ from neutron captures on \hydrogen\ escape the \lsv\ without depositing energy.
The maximum possible rejection power of the neutron capture channel alone, if every neutron capture were detected, with a 5\% concentration of \tmb, would be then $\sim99.3$\%.
The remaining inefficiency in the neutron capture rejection power comes from the tails in the neutron capture time distribution that extend outside of the acquisition window, which happens to $\sim0.2\%$ of neutrons that thermalize in the \lsv. This inefficiency brings the neutron detection efficiency by the neutron capture channel alone down to $\geq\,99.1\%$.

Estimates of the detection efficiency of the neutron thermalization signal are difficult to make, since we have not yet measured the strength of ionization quenching of proton and carbon recoils in the liquid scintillator. While we were able to measure the capture signal with the \AmBe\ source, the high rate of \grs\ in coincidence with the neutron that the source produces makes it difficult to measure reliably the prompt thermalization signal. A calibration campaign using an \AmC\ source with a much lower \gr-rate is planned. In the meantime, however, \geant\ simulations assuming the quenching model presented in~\cite{Hong:2002fn} were performed. With the current 1 PE threshold for vetoing the thermalization signal, these simulations indicate that the total neutron vetoing efficiency from both the thermalization and capture signals is well above the goal of $\geq\,99.5\%$. The total acceptance loss from both of these cuts is expected to be $\leq$16\%.

\section{Conclusions and outlook}
\label{sec:conclusion}
The construction and commissioning of the veto system of \dsf\ is completed.
Early operation shows that the system is on progress to meet the design performance goals.
\dsf\ is currently taking data with underground argon in WIMP search mode, with a neutron veto capable of ensuring a few years of background-free running.

The performance of the \dsf\ veto shows that the boron-loaded liquid scintillator technology provides an efficient way for tagging neutrons in WIMP search experiments. Boron-loaded liquid scintillator vetoes are a valuable alternative to gadolinium-loaded liquid scintillator vetoes, provided that the radioactive background contamination in the detector is low enough and that the light yield is sufficiently high.
In this case, a boron-loaded veto like the one presented here can attain a very high efficiency, since the \lith\ and $\alpha$ from the neutron capture on \borten\ are short ranged and cannot escape the detector.

\acknowledgments

The DarkSide-50 Collaboration would like to thank LNGS laboratory and its staff for invaluable  technical and logistical  support.  This report is based upon work supported by the US NSF (Grants PHY-0919363, {PHY}-{1004072}, {PHY}-{1004054}, {PHY}-{1242585}, {PHY}-{1314483}, {PHY}-{1314507} and associated collaborative grants; Grants {PHY}-{1211308} and {PHY}-{1455351}), the Italian Istituto Nazionale di Fisica Nucleare (INFN), the US DOE (Contract Nos.~{DE}-{FG02-91ER40671} and {DE}-{AC02-07CH11359}), and the Polish NCN (Grant {UMO}-{2012/05/E/ST2/02333}). We thank the staff of the Fermilab
Particle Physics, Scientific and Core Computing Divisions for their support.  We acknowledge the financial support from the UnivEarthS Labex program of Sorbonne Paris Cit\'e ({ANR}-{10-LABX-0023} and {ANR}-{11-IDEX-0005-02}) and from the S\~ao Paulo Research Foundation (FAPESP).

\bibliographystyle{veto-description}
\bibliography{veto-description.bib}

\end{document}